На правах рукописи

**ЗЫРЯНОВ Борис Анатольевич**

# РАЗВИТИЕ СИСТЕМЫ КОМПЛЕКСНОЙ ОБРАБОТКИ СИГНАЛОВ

**Специальность 05.25.07 – Исследования в области проектов и программ**

Диссертация в виде научного доклада на соискание ученой степени
доктора технических наук

Научный консультант: - действительный член РАЕН,
доктор техн. наук, профессор
**Гольдштейн Сергей Людвигович**

Екатеринбург - 2010

Работа выполнена в УПИ им.С.М.Кирова, ЗАО «Микрон», Свердловском региональном общественном фонде «Инновационные технологии», Региональном Уральском отделении Академии инженерных наук им. А.М.Прохорова, НП «Уральский межакадемический союз», г.Екатеринбург.

Официальные оппоненты:

доктор технических наук Зобнин Борис Борисович,

доктор технических наук Токманцев Валерий Иванович,

доктор технических наук Цепелев Владимир Степанович.

Защита состоится « 17 » июня 2010 года в 15-00 на заседании Диссертационного Совета Д 098.07 РСО ММС 096 по адресу: 620077, г.Екатеринбург, ул. Володарского , 4, НИИЦветмет / УМС.

Доклад разослан «_14_»_мая_ 2010 г.

С диссертацией в виде научного доклада можно ознакомиться в библиотеке ГОУ ВПО УГТУ-УПИ.

Ученый секретарь
диссертационного совета,
проф., к.ф.-м.н. В.И.Рогович

## СПИСОК СОКРАЩЕНИЙ

КОС – комплексная обработка сигналов;
СНЧ – сверхнизкочастотный (диапазон, -ая шумометрия);
СЧМС – сложная человеко-машинная система;
ГС – генеральная совокупность;
БПФ – быстрое преобразование Фурье;
БПУА – быстрое преобразование Уолша-Адамара;
СПМ – спектральная плотность мощности;
ФК – функции когерентности;
ФЧК – функции частной когерентности;
ФМК – функции множественной когерентности;
МНК – метод наименьших квадратов;
МВВКП – метод вариационно-взвешенных квадратических приближений;
РО – распознавание образов;
МО – минимизация описаний;
ПО – программное обеспечение;
ППТ – процесс производственного травматизма;
ПТ – производственный травматизм;
НС – несчастный случай;
АЭС – атомная электростанция;
ЯЭУ – ядерная энергетическая установка;
АСУ – автоматизированная система управления;
АСУ П – автоматизированная система управления производством;
АСУ ТП – автоматизированная система управления технологическими процессами;
ТП – технологические параметры;
УГТУ-УПИ – Уральский государственный технический университет – Уральский политехнический институт;
СРОФ ИТ – Свердловский региональный общественный фонд «Инновационные технологии»;
УРО АИН – Уральское региональное отделение Академии инженерных наук им. А.М.Прохорова;
I.N.A.I.L. – Национальный институт анализа производственного травматизма Италии.

## ОБЩАЯ ХАРАКТЕРИСТИКА РАБОТЫ

**Актуальность исследования.** В соответствии с классической монографией Ж.Макса [7], «невозможно дать сжатое и приемлемое для всех определение обработки сигналов. … необходимость в обработке сигналов возникает каждый раз, когда нужно отделить передаваемое сообщение от искажающего его шума.».

Традиционное представление о комплексной обработке сигналов (КОС) заключается в получении и последующей обработке в единой схеме трех и более сигналов с целью получения более полной информации об объекте исследования.

Вопросам комплексной обработки однотипных сигналов в высокочастотном диапазоне (акустика, радиоволны) с целью локации их источников посвящены работы многих отечественных и зарубежных ученых, среди которых А.И.Могильнер, В.А.Кривцов, Р.Ф.Масагутов, H.Nishihara, В.М.Соколов, T.N.Claytor, D.A.Greene, J.Maloyrh, В.В.Шемякин, К.Н.Проскуряков, А.Ю.Шатилов, М.С.Ярлыков и др.

В более сложных случаях, требующих получения информации о состоянии сложной технической системы, необходимо рассматривать в едином концептуальном пространстве разнотипные процессы, такие, например, как флюктуации физических величин при производстве замеров, шумы технологических параметров оборудования или ППТ. Необходимость комплексной обработки трех и более типов сигналов различной физической природы в постановочном плане упоминается в классических трудах Дж. Бендата и А. Пирсола [31].

Диссертационное исследование посвящено развитию системы КОС в направлении ее интеграции с распознаванием образов на основных этапах обработки, начиная с выбора частотного диапазона и комплексов процессов, как наиболее информативных выборок из генеральной совокупности.

**Координация исследования** проводилась в рамках научно-исследовательских программ Министерств среднего машиностроения (1977-1983г.г.), тяжелого и транспортного машиностроения СССР (1985-1988 г.г.),

Белоярской АЭС им.И.В.Курчатова (1982-1984 г.г.), ВЦНИИОТ ВЦСПС (1986-2004 г.г.), СРОФ ИТ (2007-2010 г.г.), УРО АИН (2005-2010 г.г.).

**Объект исследования –** комплексная обработка сигналов (КОС).

**Предмет исследования** – развитие КОС разнотипных процессов и ее интеграции с РО, как технического инструментария анализа состояния сложных систем.

**Цели и задачи исследования**

Глобальная цель исследования – развитая система КОС.

Локальная цель 1 – получение нового знания в виде пакета прототипов и моделей системы КОС.

Локальная цель 2 – инженерная реализация и внедрение системы КОС для разных видов задач, разработка программного обеспечения.

Задачи исследования:

- подготовка литературно-аналитического обзора ретроспективы и современного состояния исследований обработки сигналов методами спектрального анализа в различных частотных диапазонах;
- составление пакетов научных и корпоративных прототипов по теме исследования;
- совершенствование методов минимизации описаний;
- разработка моделей применения КОС для изучения невязки, шумов технологических параметров ЯЭУ и оборудования АЭС;
- изучение ППТ методами КОС;
- проведение сравнительного анализа результатов обработки сигналов процесса ПТ России и Италии;
- внедрение полученных результатов и моделей в научных исследованиях физических процессов, при разработке АСУ ТП АЭС, в учебном процессе.

**Эмпирической базой** исследования послужили материалы опубликованных научных разработок, выполнявшихся с участием автора в СФ НИКИЭТ, УГТУ-УПИ, Белоярской АЭС; база данных ППТ, сформированная в отечественных и зарубежных командировках, на основе анкет НС по России и банка данных Италии, переданного автору I.N.A.I.L.

**Методы исследования:** спектральный анализ, распознавание образов, минимизация описаний, определение статистических характеристик и функций случайных процессов, моделирование.

**Достоверность результатов** подтверждена машинным моделированием и проведением полного комплекса стандартных математических процедур исследования статистической достоверности, тестами эргодичности и стационарности; совпадением результатов обработки данных, различающихся хронологически, полученных на нескольких участках технологического оборудования, а в отношении НС - регистрируемых на разных предприятиях, и в странах с заметно отличающимися условиями труда. Обусловлена апробацией в научных изданиях, семинарах и конференциях основных результатов исследования и их практическим внедрением в научных исследованиях и на промышленных предприятиях.

**Научная новизна:**

1. На основе структурного анализа литературной информации впервые сформирован пакет научных прототипов комплексной обработки сигналов (КОС), отличающийся 4-х-ранговой структурой.
2. Разработаны проекты и программы развития КОС, отличающиеся усовершенствованием методов обработки и их применения для анализа разных типов процессов на основе интеграции с распознаванием образов (РО).
3. Сформулирована общая информационная модель обработки сигналов по комплексам наиболее информативных процессов – выборкам из генеральной совокупности, определяемым с применением РО.
4. Обоснована необходимость использования СНЧ диапазона процессов, т.е. области частот от нано- до десятых долей Гц, для КОС сложных технических систем.
5. Предложен новый метод минимизации описаний (МО) на основе ранжирования коэффициентов разделяющих функций, отличающийся большей устойчивостью результатов распознавания и гарантированным разделением исходной совокупности на классы при решении сложных задач РО.

6. Развита система классификации признаков на дифференциальные, интегральные и интегро-дифференциальные, позволившая расширить исходное признаковое пространство и применить ранее не использовавшиеся, информативные признаки, такие, как коэффициенты разделяющих функций, меры сходства, значения ФЧК и ФМК, биспектры, распределение переходов знака невязки.

7. Разработан новый метод МО путем интеграции минимизации описаний по разрешающей способности и идеи случайного поиска с адаптацией, отличающийся большей универсальностью, поскольку при изменении описания результаты распознавания становятся более устойчивыми.

8. Разработанная информационная модель КОС в применении к анализу процессов на АЭС в СНЧ диапазоне позволила получить диагностические модели обнаружения кипения в активной зоне ядерного реактора, «образа» шумов исправного парогенерирующего, насосного и конденсаторного оборудования АЭС с использованием штатной измерительной аппаратуры энергоблока.

9. Развито представление о невязке, как выходном процессе информационной модели КОС, содержащем информацию для уточнения представлений о детерминированной основе изучаемого физического процесса.

10. Применение информационной модели КОС позволило развить представление о детерминированных основах процессов в задачах изучения низкотемпературной плазмы, а также проектирования материалов биологической защиты (пропускание нейтронного потока, газовыделение).

11. На основе применения информационной модели КОС открыта зависимость между ранее считавшимися не связанными категориями, такими, как ППТ и циркадианные ритмы, а в более общем плане, установлен детерминированный, полигармонический характер ППТ, с определяющим воздействием циркадианных составляющих и второстепенным – от антропогенных факторов. Доказано отсутствие влияния на ППТ околомесячных биоритмов, широко применяемых на предприятиях ряда стран. Разработана автоколебательная модель ППТ, предложены методы его рандомизации.

**Практическая значимость:**

- развитая система КОС в исследованиях невязки позволила уточнить физические измерения и теоретические представления о процессах, дать более точные рекомендации для проектирования материалов биологической защиты от ионизирующих излучений в СФ НИКИЭТ;
- результаты применения КОС в исследованиях комплексов шумов и распознавания режимов работы внедрены на исследовательской ЯЭУ в СФ НИКИЭТ; соответствующие методики диагностики технологического оборудования использованы на Белоярской и Курской АЭС и Молдавской ГРЭС, рекомендованы для внедрения Минэнерго СССР;
- результаты применения развитой КОС в исследованиях ППТ внедрены на Ясногорском машиностроительном заводе (г.Ясногорск) и ВЦНИИОТ ВЦСПС (г.Москва);
- намечены пути дальнейших исследований ППТ и начальные меры по совершенствованию системы инструктажей и снижению уровня травматизма на производстве;
- результаты исследований и разработанное программное обеспечение внедрены в учебном процессе ряда ВУЗов, в частности, УГТУ-УПИ, ОПИ, УрГУ, а также в Институте электрофизики УРО РАН и СФ ИПК Минлесбумпрома СССР;
- получены соответствующие акты внедрения результатов работы в вышеуказанных организациях и министерствах.

**Положения, выносимые на защиту:**

1. В комплексной обработке сигналов (КОС) отсутствуют системотехнические исследования, направленные на изучение структуры данного научного направления и его совершенствование.

   В диссертации разработан 4-х-ранговый пакет прототипов, приведена их критика, показаны алгоритм и схема функционирования системы КОС.

2. В исследованиях, связанных с комплексной обработкой сигналов разнотипных процессов, выбор реализаций осуществляется на основе априорной или экспертной информации с учетом целей исследования, формализованный подход отсутствует.

В диссертационном исследовании сформулирован более общий, алгоритмизуемый подход на основе развитой информационной модели, использующей комплексы наиболее информативных в смысле целевой функции процессов – выборок из генеральной совокупности.

3. СНЧ диапазон остается недооцененным в диагностическом плане.

В диссертации (на исследованных примерах) показаны диагностические преимущества СНЧ диапазона: высокая «проникающая» способность, получение более общей информации об исследуемом объекте или процессе, и, что особенно важно, возможность формирования в этом диапазоне комплексов процессов из широкой, заведомо избыточной выборки из генеральной совокупности без привязки к физической природе и частоте процесса локальной задачи.

4. В исследованиях, посвященных поиску детерминированной основы физических процессов, принято представление о невязке, как проявлении помех и несовершенства измерительных систем. Соответственно, применяется цифровая или аналоговая фильтрация для приближения аппроксимационной зависимости к полученным экспериментальным данным.

В диссертации развито представление о невязке, как выходном процессе информационной модели КОС, содержащем данные для уточнения искомой детерминированной основы.

5. Распознавание образов (РО) в исследованиях, связанных с технической диагностикой и прогнозированием, используется только на заключительном этапе для формирования условий принятия решений.

В диссертационном исследовании сформулирована необходимость применения системы РО на начальном этапе исследования для формализованного выбора наиболее информативных комплексов процессов.

6. Существующие методы минимизации описаний (МО) в РО имеют узкую область применения и при распознавании в сложных задачах недостаточны.

В диссертационном исследовании разработан более универсальный метод МО путем использования идеи случайного поиска с адаптацией в развитие метода МО по разрешающей способности.

7. Недостатком существующих методов МО является неустойчивость результатов распознавания при изменениях признакового пространства. Особенно это проявляется при частичном «перекрытии» признаков, когда устойчивое разделение на классы не достигается.

В диссертационном исследовании этот недостаток устранен путем развития метода МО весового ранжирования признаков, принятием в качестве признаков коэффициентов разделяющих функций.

8. В исследованиях по охране труда и ПТ принято феноменологическое представление о травматизме, как совокупности случайных многопараметрических величин.

В диссертации, на основе адаптации КОС к исследованиям ППТ в сверхнизкочастотном диапазоне показано, что травматизм является детерминированным, полигармоническим процессом с определяющим воздействием циркадианных составляющих и второстепенным – антропогенных факторов.

9. В мировой практике воздействия на ППТ с целью его уменьшения широко используется теория влияния на ППТ биоритмов с гармониками, близкими к календарному месяцу.

В диссертации доказано, что такое представление подлежит корректировке, а ППТ не содержит значимых гармоник такой частоты.

**Апробация результатов исследования.** Основные результаты настоящего диссертационного исследования в период 1979-2009 г.г. были представлены на следующих научных конференциях и семинарах:

- V зональная конф. «Применение радионуклидов и ионизирующих излучений в научных исследованиях и народном хозяйстве Урала» - Свердловск: УПИ, 1979;
- Всесоюзный н-техн. семинар «Применение вычислительной техники для решения краевых задач в экологии» - Свердловск: УПИ, 1981;
- 7 и 8 Всесоюзных конф. по опыту разработки и эксплуатации АСУ – Свердловск: УПИ, 1982 и 1983;

- I Межотраслевой семинар «Методы и программы расчета ядерных реакторов» - М.: ИАЭ им.И.В.Курчатова, 1983;
- Республиканской н-техн. конф. «Повышение эффективности работы конденсационных установок и систем охлаждения циркуляционной воды тепловых и атомных электростанций» - Киев: ИПМЭ, 1983;
- I и II областных н-техн. конф. «Актуальные проблемы атомной науки и техники» - Свердловск: Свердловский обл. совет НТО, НТО Э и ЭП, 1984;
- IV Республиканской н-техн. конф. «Современные проблемы энергетики» - Киев: ИПМЭ, 1985;
- Конф. «Некоторые актуальные проблемы создания и эксплуатации турбинного оборудования» - Свердловск: УПИ, 1986;
- II научный семинар по проблемам охраны труда и окружающей среды – М.:МАТИ, 1987;
- Всесоюзная н-практ. конф. по проблемам охраны труда в условиях ускорения научно-технического прогресса – М.: ВЦНИИОТ ВЦСПС, 1988;
- Н-техн. семинар «Машиностроение. 21 век: робототехника и нанотехнологии» в рамках IV Евро-Азиатской промышленной выставки – Екатеринбург: СРОФ ИТ, 2008;
- Коллегия СРОФ ИТ с участием УРО АИН – Екатеринбург, 2008;
- Семинары кафедры «Охрана труда» УПИ, Департамента электроники Миланского политехнического института, I.N.A.I.L., НТО Ясногорского машиностроительного завода, ВЦНИИОТ ВЦСПС, НТО ЗАО «Микрон» - Милан, Рим, Ясногорск, Москва, Екатеринбург, 1989 – 2009.

**Публикации:** по материалам диссертации имеется более 50 работ, в том числе учебное пособие, 4 монографии, авторское свидетельство на изобретение. Приоритетные публикации изданы в ведущих отечественных отраслевых журналах, а также в Италии и США.

**Личный вклад автора** состоит в определении стратегии и тактики диссертационного исследования с обработкой и интерпретацией результатов; формулировке феноменологической концепции КОС с использованием РО для выбора комплексов процессов из генеральной совокупности; разработке нового и усовершенствованию известного методов МО; усовершенствовании системы формирования описаний, включающей ранее не применявшиеся информативные признаки; разработке соответствующего ПО; применении КОС к исследованию разных типов процессов – флюктуациям измерений физических зависимостей и констант, шумам технологических параметров ЯЭУ и оборудования АЭС, а также ПТ, интерпретации полученных результатов; формулировке автоколебательной модели ППТ.

**Структура диссертационного исследования** представлена на рис.1.

Подпроекты: 1.1.1 – методы и примеры применения КОС, 1.1.2 – исследования КОС в СНЧ диапазоне, 1.1.3 – методы распознавания образов; 1.2.1 – аналоги, 1.2.2 – пакет прототипов; 2.1.1 – схема прототипов нулевого и первого рангов, 2.1.2 – схема прототипов второго ранга; 2.2.1 – алгоритм КОС по прототипу нулевого ранга, 2.2.2 – схема функционирования КОС по прототипу нулевого ранга; 2.3.1 – информационная модель КОС различной физической природы, 2.3.2 – пакет ПО для КОС; 3.1.1 – выбор диагностических процессов, 3.1.2 – формирование и математическая обработка сигналов, 3.1.3 – вычисление признаков; 3.3.1 – формирование признакового пространства, 3.3.2 – минимизация описаний; 4.1.1 – моделирование сигналов в СНЧ диапазоне, 4.1.2 – минимизация описаний; 4.2.1 – формирование выборок комплексов процессов, 4.2.2 – формирование признакового пространства, 4.2.3 – минимизация описаний и выбор комплексов процессов; 5.1.1 – пропускание через вещество нейтронного потока, 5.1.2 – газовыделение из материалов биологической защиты, 5.1.3 – зондовые характеристики низкотемпературной плазмы; 5.3.1 – флюктуации ТП в технологических трубопроводах, 5.3.2 – диагностика и прогнозирование состояния конденсационных установок; 5.4.1 – алгоритм адаптации КОС к исследованиям ППТ, 5.4.2 – обработка сигналов ППТ России и Италии, 5.4.3 – автоколебательная модель ППТ, 5.4.4 - сравнение результатов обработки сигналов ППТ Италии и России.

Социальный заказ ↓ | Логика, состояние проблемы ↓

**Программа 1** – Анализ современного состояния КОС

| **Проект 1.1 –** Литературно-аналитический обзор | | |
|---|---|---|
| **Подпроекты** | | |
| 1.1.1 | 1.1.2 | 1.1.3 |

| **Проект 1.2 -** Прототипирование | |
|---|---|
| **Подпроекты** | |
| 1.2.1 | 1.2.2 |

**Программа 2** – Моделирование прототипов и предлагаемых решений по КОС

| **Проект 2.1** Схематичес-кое представление системы прототипов и предлагаемых решений | |
|---|---|
| **Подпроекты** | |
| 2.1.1 | 2.1.2 |

| **Проект 2.2 –** Алгоритмическое моделирование КОС, ее систем и подсистем | |
|---|---|
| **Подпроекты** | |
| 2.2.1 | 2.2.2 |

| **Проект 2.3 –** Информационная модель КОС и программное обеспечение | |
|---|---|
| **Подпроекты** | |
| 2.3.1 | 2.3.2 |

**Программа 3 –** Методы комплексной обработки сигналов различной физической природы

| **Проект 3.1 –** Техни-ческая диагностика | | |
|---|---|---|
| **Подпроекты** | | |
| 3.1.1 | 3.1.2 | 3.1.3 |

**Проект 3.2 –** Техническое прогнозирование

| **Проект 3.3 –** Распознавание образов | |
|---|---|
| **Подпроекты** | |
| 3.3.1 | 3.3.2 |

**Программа 4 –** Интеграция КОС с распознаванием образов

| **Проект 4.1 –** Выбор частотного диапазона | |
|---|---|
| **Подпроекты** | |
| 4.1.1 | 4.1.2 |

| **Проект 4.2 –** Минимизация описаний комплексов процессов | | |
|---|---|---|
| **Подпроекты** | | |
| 4.2.1 | 4.2.2 | 4.2.3 |

**Программа 5 –** Примеры применения КОС

| **Проект 5.1 –** Моделирование невязки | | |
|---|---|---|
| **Подпроекты** | | |
| 5.1.1 | 5.1.2 | 5.1.3 |

**Проект 5.2 –** Комплексная обработка сигналов ТП ЯЭУ

| **Проект 5.3 -**КОС оборудования АЭС | |
|---|---|
| **Подпроекты** | |
| 5.3.1 | 5.3.2 |

| **Проект 5.4 –** Адаптация КОС к исследованиям процесса ПТ | | | |
|---|---|---|---|
| **Подпроекты** | | | |
| 5.4.1 | 5.4.2 | 5.4.3 | 5.4.4 |

**Программа 6 -** Внедрение моделей и результатов применения КОС

**Проект 6.1 –** внедрение в учебном процессе

**Проект 6.2 –** КОС в физических измерениях

**Проект 6.3 –** внедрение в АСУ ТП АЭС

**Проект 6.4 –** использование в АСУ П предприятия

↓ Новые знания | ↓ Выполненные заказы

Рис.1. Структура программ и проектов по теме диссертационного исследования

# ОСНОВНОЕ СОДЕРЖАНИЕ РАБОТЫ

## ПРОГРАММА 1 – АНАЛИЗ СОВРЕМЕННОГО СОСТОЯНИЯ КОС

Программа включает 2 проекта с 5 подпроектами.

### Проект 1.1 Литературно-аналитический обзор

В составе проекта выполнено 3 подпроекта.

Просмотрено 400 библиографических источников за период с 1964 по 2010 г.г., более 1000 адресов Интернет, опрошено 35 экспертов.

**Подпроект 1.1.1** Методы и примеры применения КОС

Анализ литературных источников, проведенный в **подпроекте 1.1.1,** показал, что большая часть работ по КОС посвящена технической диагностике и исследованиям процессов на оборудовании АЭС и ТП ЯЭУ. Это объясняется объективной необходимостью диагностики таких систем по косвенным характеристикам, ввиду специфики объекта исследования.

Изучение возможностей обработки сигналов основывается, главным образом, на регистрации однотипных ТП (нейтронного потока, температуры, давления). В ряде работ исследуются вопросы парного взаимодействия флюктуаций типа «нейтронный поток – давление, - расход, - паросодержание», основанные на эффекте увеличения корреляции этих параметров при диагностике кипения в активной зоне, а также пар «вибрации – давление», однотипных «мощность (нейтронный поток) – мощность» и других с целью получения передаточных функций и времени задержки сигналов. Практически не исследованы вопросы взаимодействия трех и более ТП различной физической природы.

Техническое прогнозирование [18)] имеет больше областей применения, однако в литературе отсутствует такой подход, как распознавание детерминированной основы путем КОС процесса невязки.

В таких приложениях КОС, как хронобиология, отсутствует развитое применение аппарата распознавания образов, а в травматологии КОС ранее не применялась.

**Подпроект 1.1.2** Исследования КОС в СНЧ диапазоне

Установлено, что частотный диапазон, в котором заключена основная часть информации, является примерно одинаковым для разных типов процессов и составляет от 0 до 50 Гц, причем зарегистрированное во многих работах максимальное значение СПМ находится в начале частотной координаты. Это означает, что пик спектрограммы был определен только одной точкой в интервале частот от нуля до величины экспериментального разрешения – 0,04 Гц [1)] , 0,5 Гц [2)] , 0,2 Гц [3)] , 0,01 Гц [4,5)] , 0,12 Гц [6)] .

Исследований, направленных на более детальное изучение СНЧ максимума не проводилось. Причиной этого является естественное желание экспериментаторов рассматривать всю область частот, содержащую основную часть интегрального уровня шума, а также то, что в результате многочисленных исследований в диапазоне 0-50 Гц появляется кажущаяся изученность диапазона, например, 0-0,05 Гц. Однако, в силу принципа неопределенности, увеличение частоты отсчетов при фиксированном, приемлемом для машинной обработки количестве экспериментальных точек, снижает разрешение и СНЧ область не проявляется. Например, для проведения измерений в диапазоне 0-50 Гц и задания СНЧ области 0-0,05 Гц СПМ ста значениями, требуется зарегистрировать и сделать преобразование Фурье 100 000 точек. Реализации такой длины в высокой степени уязвимы для помех. Практически для СНЧ исследований необходимо получение реализаций до нескольких тысяч точек с большим интервалом времени между отсчетами (от $10^1$ до $10^4$ с), что снижает верхнюю граничную частоту.

В работе Ж.Макса [7)] сделан принципиально важный вывод о том, что наиболее содержательная часть спектра в исследованиях шумов реакторов заключена в низкочастотной области с верхней граничной частотой 0,002 - 2 Гц и ниже. Ж.Макс отмечает актуальность изучения этой области частотного диапазона и выдвигает еще одну причину, объясняющую отсутствие таких работ, связанную с тем, что накопление отсчетов, необходимое для проявления низкочастотных составляющих СПМ, требует значительного времени, вследствие чего становится невозможным создание диагностических систем режима он-лайн.

Тем не менее, в большинстве случаев инерционные штатные информационно-вычислительные системы обеспечивают получение данных в СНЧ диапазоне. Сверхнизкие частоты отражают глобальные изменения процессов, имеющие, как правило, большую мощность, чем локальные. Их «проникающая» способность выше, что немаловажно для целей комплексной обработки сигналов (КОС) и полномасштабной диагностики сложных технических систем.

На рис.2 приведена разработанная диаграмма частотных интервалов процессов, вызываемых различными физическими причинами (**подпроект 1.1.2)**. Диаграмма не включает данные о физических процессах, рассмотренных далее в проекте 5.1, поскольку модели КОС в таких приложениях не требуют привязки к частотному интервалу.

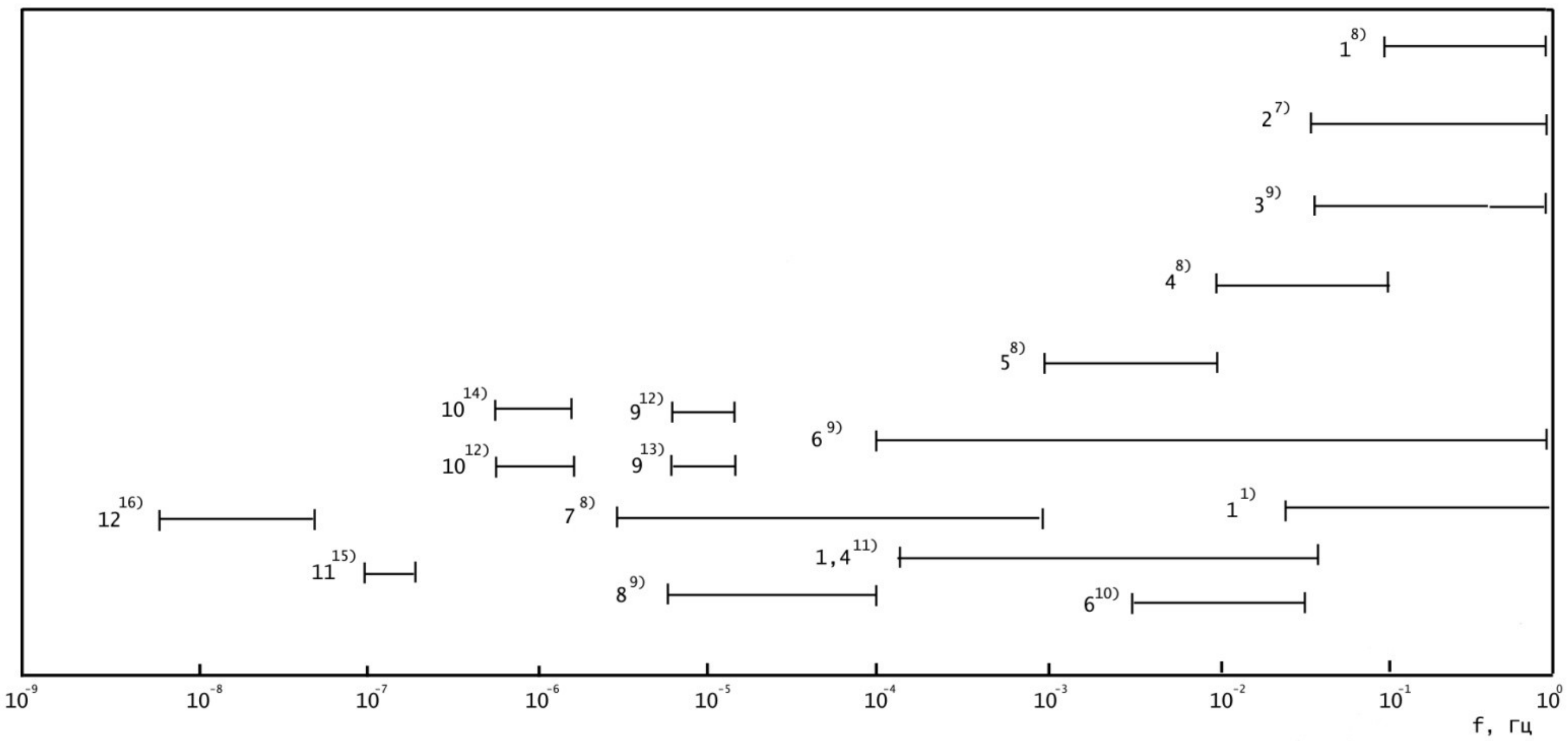

Рис.2. СНЧ диапазон: частотные интервалы процессов различной природы.
1- циркуляция теплоносителя на АЭС [1, 8, 11], 2 – термодинамические явления, например, флюктуации тепловой мощности [7], 3 – гидравлические процессы [9], 4 – автоматическое регулирование [8, 11], 5 – охлаждение внутрикорпусных устройств [8], 6 – эффекты реактивности и замедление нейтронов [9, 10], 7 – ксеноновое отравление и охлаждение корпуса [8], 8 – эксплуатационные факторы, например, колебания поля энерговыделения [9], 9 – атмосферные воздействия и циркадианные ритмы [12, 13], 10 – недельные процессы [12, 14], 11 – лунные биоритмы [15], 12 – сезонные атмосферные воздействия, изменение реактивности ЯЭУ во время кампании и после перегрузок (отравление, выгорание, зашлакование) [16].

В **подпроекте 1.1.3** рассмотрены известные в литературе, в частности [17,19], основные методы распознавания образов. Это, прежде всего, формирование признакового пространства, минимизация описаний (МО), построение разделяющих функций.

**Проект 1.2 Прототипирование**

В составе проекта выполнено 2 подпроекта.

В **подпроекте 1.2.1** на основе литературного обзора выявлены аналоги решений: КОС в радио – и акустическом диапазонах, низкочастотная шумометрия ТП, а **подпроекте 1.2.2** сформирован пакет прототипов (табл.1).

Таблица 1. Пакет прототипов и их критика

| Прототипы | | | Критика прототипов |
|---|---|---|---|
| Ранг | Название | Источник | |
| 1 | 2 | 3 | 4 |
| 0 | Обобщенный прототип: Комплексная обработка сигналов | 7,29,31,34, 41,42 | Нет единого модельного комплексного представления взаимозависимости трех и более процессов разной физической природы, отсутствует распознавание образов на этапе формирования многопараметрической модели, общая модельная направленность на решение локальных задач; не ставится задача и не достигается функция получения максимума диагностической информации, структурная и функциональная неполнота. |
| 1 | 1.1. Система методов обработки сигналов | 7,10,29 | Распознавание образов в задачах технической диагностики используется недостаточно, только на заключительных этапах, связанных с принятием решений. Задача определения с помощью распознавания образов диагностических и прогностических комплексов процессов не ставится. Функциональная неполнота. |
| | 1.2. Система примеров применения | 20,22-24,35,39,40 | Модели с использованием комплексов процессов ограничиваются рассмотрением двух – трех однотипных процессов. Не рассматриваются взаимодействие и взаимное влияние трех и более процессов различной физической природы. Невязка в практических приложениях не рассматривается, как объект модельного исследования с применением распознавания образов для получения данных о детерминированной основе процессов. Функциональная неполнота. |
| | 1.3. Система интеграции с распознаванием образов | 17,19,26 | Не ставится задача формализованного определения частотного диапазона и комплексов процессов с использованием методов распознавания образов. Функциональная неполнота. |
| | 2.1. Подсистема технической диагностики | 8-10,37 | Не формализован выбор диагностических процессов. Комплексы процессов не используются для описания функционирования СЧМС в целом, рассматриваются только |
| 2 | | | локальные задачи. Функциональная неполнота. |
| | 2.2. Подсистема технического прогнозирования | 18,28,35 | Не ставится задача распознавания детерминированной основы процессов для прогнозирования. Моделирование проводится в статике, без учета эксплуатационных процессов. Функциональная неполнота. |
| | 2.3. Подсистема интеграции с распознаванием образов | 17,19,26 | Признаковое пространство и методы МО недостаточны для решения сложных задач. Функциональная неполнота. |
| | 2.4. Подсистема реакторной шумометрии | 1-6,8,9, 25,35 | Не проявлен СНЧ диапазон, соответственно решаются локальные задачи. Функциональная неполнота. |
| | 2.5. Подсистема материаловедения | 16,39 | Не ставится задача распознавания детерминированной основы процессов. Функциональная неполнота. |
| | 2.6. Подсистема геофизики и сейсмологии | 29,40 | Не ставится задача распознавания детерминированной основы процессов. Функциональная неполнота. |
| | 2.7. Подсистема травматологии | 12-15,30,33 | Нет рассмотрения ПТ как процесса, не достигается функция получения информации о детерминированной основе процесса, функциональная недостаточность. |
| | 2.8. Подсистема выбора частотного диапазона с использованием распознавания образов | 7 | Не ставится задача формализованного выбора частотного диапазона с использованием методов распознавания образов. Функциональная неполнота. |
| | 2.9. Подсистема минимизации описаний комплексов процессов | 26 | Признаковое пространство и предлагаемые методы МО не позволяет решать сложные задачи выбора комплексов процессов. Функциональная неполнота. |
| 3 | 3.1. Блок выбора диагностических процессов | 10,37 | Не формализован выбор диагностических процессов и их комплексов. Функциональная неполнота. |
| | 3.2. Блок формирования и математической обработки сигналов | 7,31,37 | Не ставится задача моделирования сигналов для определения частотного диапазона. Функциональная неполнота. |
| | 3.3. Блок вычисления признаков | 17,26 | Предлагаемых признаков недостаточно для решения сложных задач. Функциональная неполнота. |
| | 3.4. Блок определения детерминированной основы процесса с использованием распознавания образов | 18 | Не ставится задача распознавания детерминированной основы с формированием признакового пространства и минимизацией описаний. Функциональная неполнота. |
| | 3.5. Блок формирования признакового пространства | 17,19 | При формировании описаний не рассматриваются функции комплексов процессов в качестве комплексных интегро-дифференциальных признаков задачи распознавания. Функциональная неполнота. |
| | 3.6. Блок минимизации описаний | 26 | Приведенные методы МО не позволяют устойчивое разделение образов в сложных задачах. Функциональная неполнота. |
| | 3.7. Блок формирования | | Не рассматривается формализованное формиро- |
| | описаний процесса | 6,8-9 | вание описаний на основе выборок комплексов реализаций из генеральной совокупности. Функциональная неполнота. |
| | 3.8. Блок диагностики состояния процессов и оборудования | 20-25 | Решаются задачи локального характера, изучаются отдельные виды процессов. Функциональная неполнота. |
| | 3.9. Блок моделирования невязки с использованием распознавания образов | 29,32,39 | Отсутствует интеграция с распознаванием образов и поиск детерминированной основы для решения задач прогнозирования. Функциональная неполнота. |
| | 3.10.Блок моделирования и прогнозирования | 29,36,40 | Не решается вопрос соответствия теоретических представлений о детерминированной основе процесса результатам обработки сигналов. Функциональная неполнота. |
| | 3.11.Блок комплексного спектрального исследования | 30 | Отсутствует. Нет рассмотрения ПТ как процесса, не достигается функция получения информации о детерминированной основе процесса, функциональная недостаточность. |
| | 3.12.Блок моделирования откликов с повышением разрешения в СНЧ диапазоне | 7 | Слишком широкий частотный интервал не дает детального проявления СНЧ пиков, нет соответствующего моделирования, функциональная неполнота. |
| 3 | 3.13. Блок минимизации описаний | 26,27 | В сложных для кластеризации задачах признаки по предлагаемым критериям не различаются, полное разделение классов не достигается; исключение любого признака меняет меры сходства с непредсказуемым результатом распознавания; увеличение размерности пространства признаков приводит к смещению положения центров рассеяния описаний, веса признаков меняются нелинейно, и первоначально информативные признаки становятся менее «весомыми»; ряд критериев введен для разделения только двух классов, основные недостатки те же. Функциональная неполнота. |
| | 3.14. Блок формирования выборок комплексов процессов из генеральной совокупности | 7,31 | Формирование выборок процессов осуществляется с помощью только априорной информации экспертного характера. Функциональная неполнота. |
| | 3.15. Блок формирования признакового пространства | 17,19,26 | Признаковое пространство является недостаточно полным для решения сложных задач выбора комплексов процессов. Необходимо расширение признакового пространства за счет комплексных интегральных и интегро-дифференциальных признаков. Функциональная неполнота. |
| | 3.16. Блок минимизации описаний и выбора комплексов процессов | 26 | Рассматриваемые методы МО не дают устойчивого разделения на классы в сложных приложениях. Функциональная неполнота. |

[1] Власов В.И., Мокрушин С.А., Радченко Р.В. Исследование теплогидравлической неустойчивости в активной зоне реактора ИВВ // Физика и техника реакторов, Л.: ЛИЯФ, 1982, с.30-45.

[2] Bauernfeind V. Vibration and pressure signals as sources of information for an on-line vibration monitoring system in PWR power plants // Nucl.Eng.and Des. – 1977.-Vol.40, №2, pp.403-420.

[3] Minelli M.R. et al. Detection of cooling anomalies in IMFBR noise analysis techniques // Annals of Nuclear Energy.-1980.-Vol.7, №11, pp.623-629.

## ПРОГРАММА 2 – МОДЕЛИРОВАНИЕ ПРОТОТИПОВ И ПРЕДЛАГАЕМЫХ РЕШЕНИЙ ПО КОС

Программа включает 3 проекта и 6 подпроектов.

**Проект 2.1 Схематическое представление системы прототипов и предлагаемых решений**

В составе проекта 2.1 выполнено 2 подпроекта. Схема прототипов и предлагаемых решений нулевого и первого рангов представлена ниже на рис.3, а второго ранга – на рис.4.

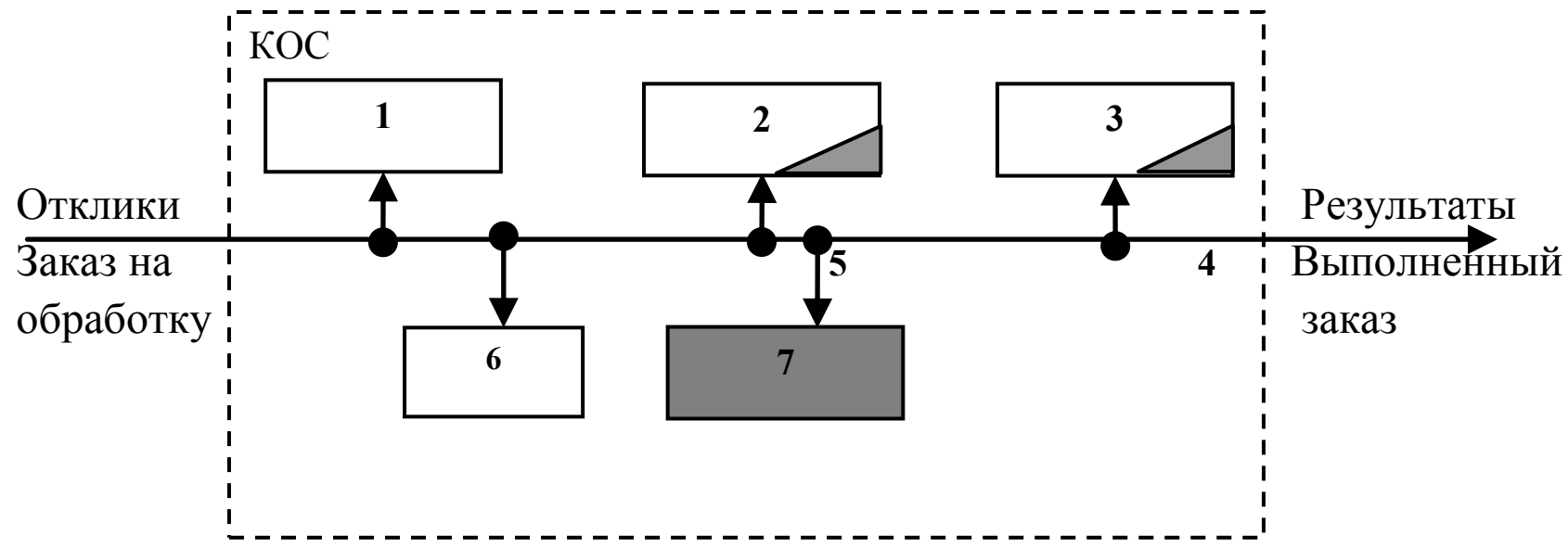

*Обобщенный прототип 0 ранга* – комплексная обработка сигналов (КОС). Системы: 1 – техники обработки сигналов, 2 – методов обработки сигналов, 3 - примеров применения, 4,5 - интерфейсы, 6 - адаптации КОС к конкретной задаче, 7 - интеграции с распознаванием образов.

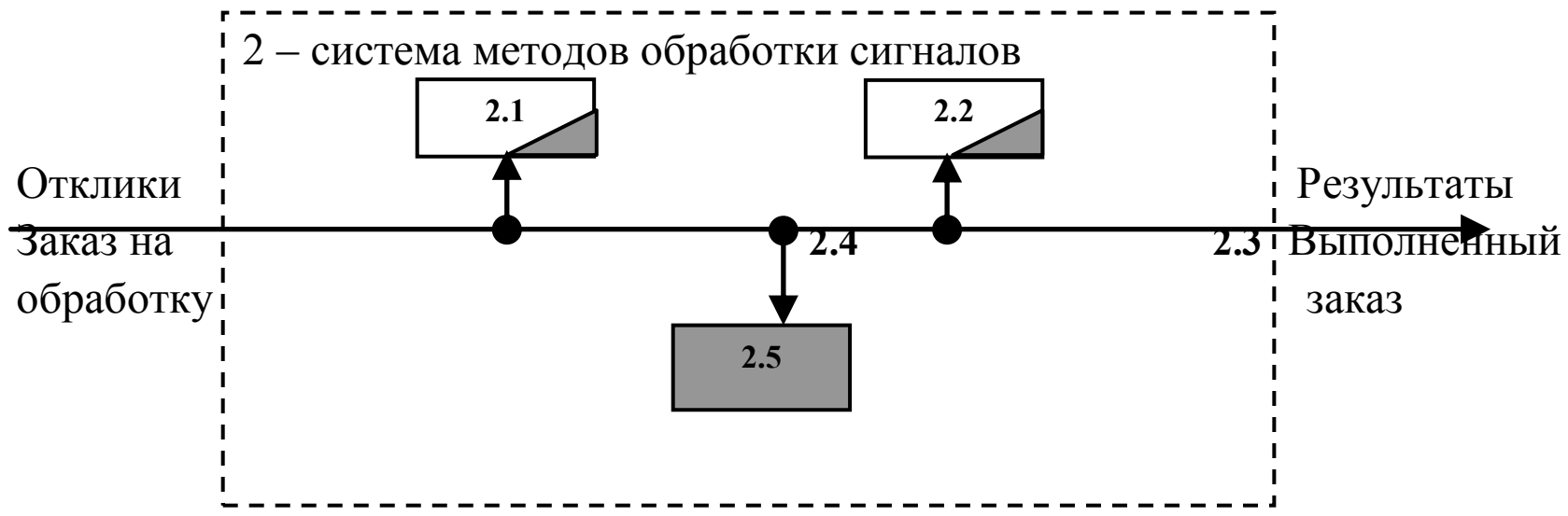

*Первый прототип 1 ранга* – система методов обработки сигналов. Подсистемы: 2.1- технической диагностики, 2.2- технического прогнозирования, 2.3, 2.4- интерфейсы, 2.5- интеграции с распознаванием образов.

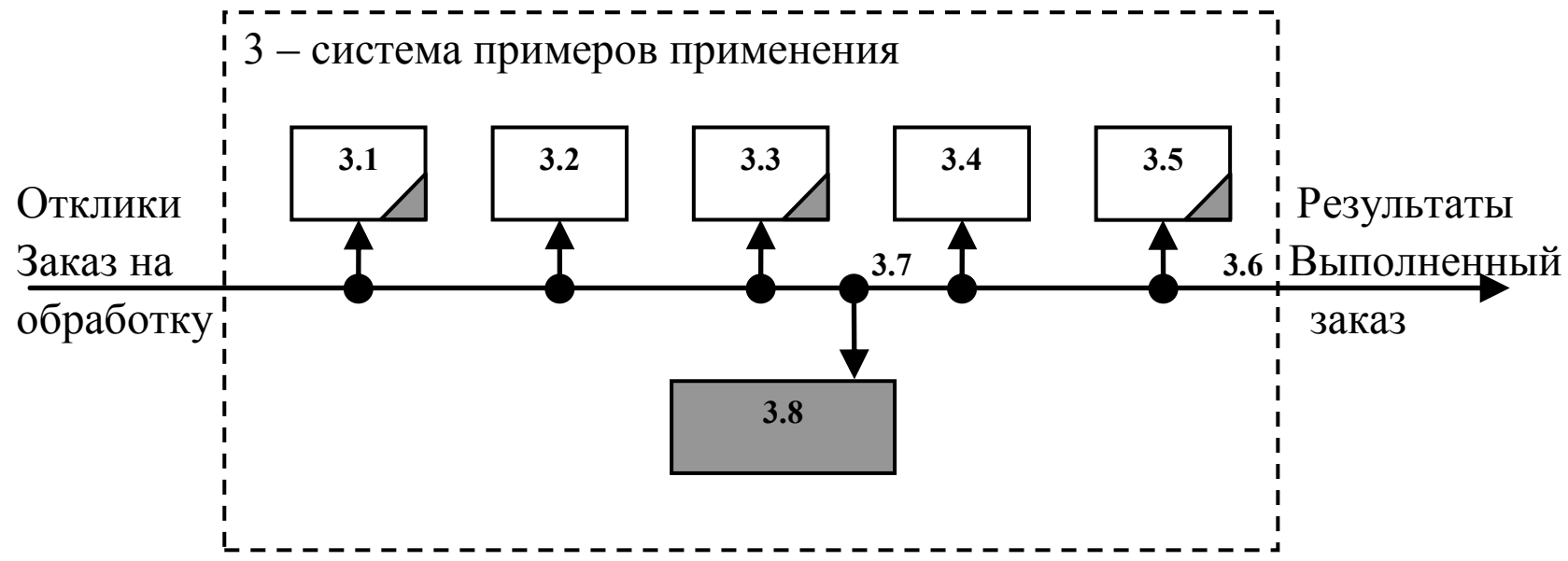

*Второй прототип 1 ранга* – система примеров применения. Подсистемы: 3.1 – реакторной шумометрии, 3.2 – астрофизики, 3.3 – материаловедения, 3.4 – биомедицины, 3.5 – геофизики и сейсмологии, 3.6, 3.7 – интерфейсы, 3.8 – травматологии.

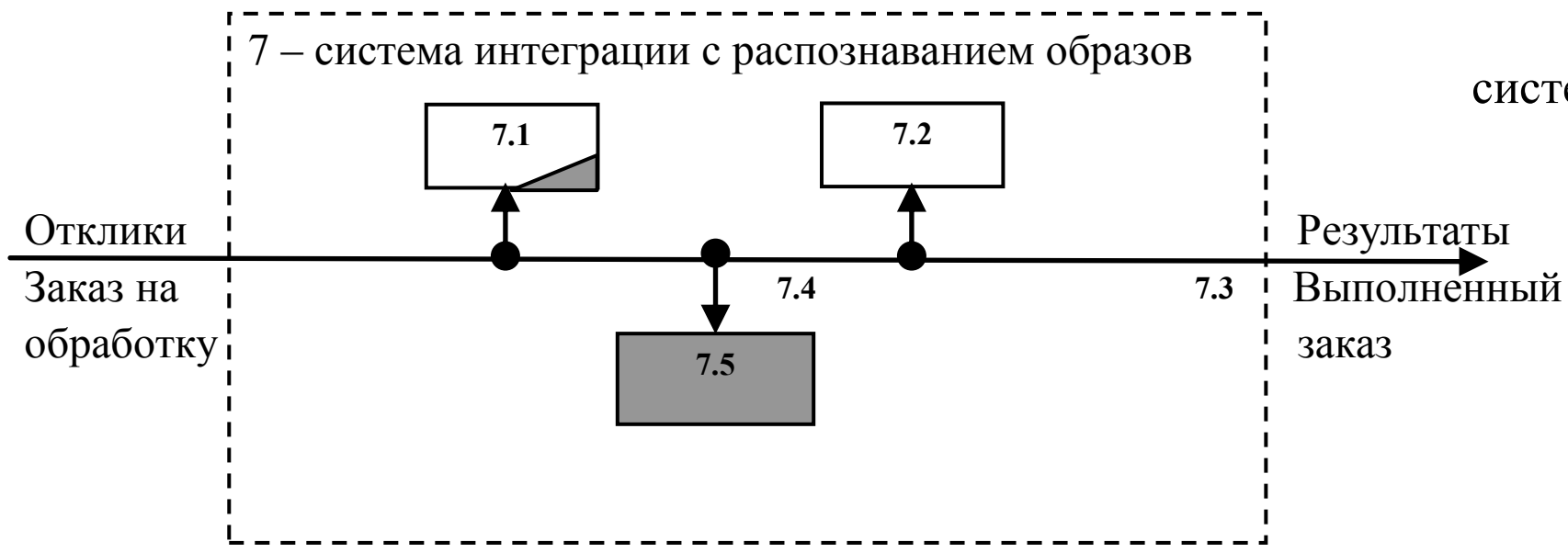

*Третий прототип 1 ранга* – система интеграции с распознаванием образов. Подсистемы: 7.1 – выбора частотного диапазона с использованием распознавания образов, 7.2 – принятия решений с помощью распознавания образов, 7.3, 7.4 – интерфейсы, 7.5 – минимизации описаний комплексов процессов.

Рис.3. Схема прототипов и предлагаемых решений нулевого и первого рангов (**подпроект 2.1.1**).

На всех рисунках в данном подпроекте и далее серый фон в уголке прямоугольников означает, что в диссертации в данном разделе разработаны частичные новые решения, а полностью затененные элементы означают, что они применены впервые или содержат преобладающую степень новизны в данном разделе работы.

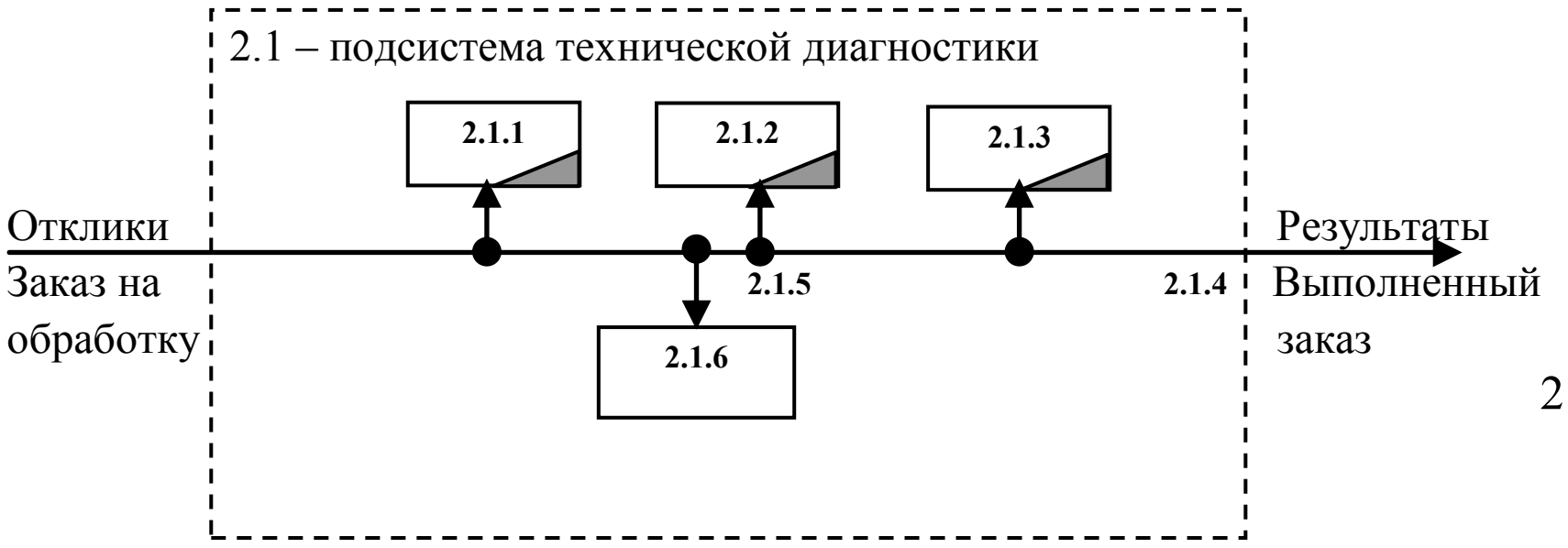

*Первый прототип 2 ранга* – подсистема технической диагностики. Блоки:

2.1.1 – выбора диагностических процессов, 2.1.2 – формирования и математической обработки сигналов, 2.1.3 – вычисления признаков, 2.1.4, 2.1.5 - интерфейсы, 2.1.6 – принятия решений с помощью распознавания образов.

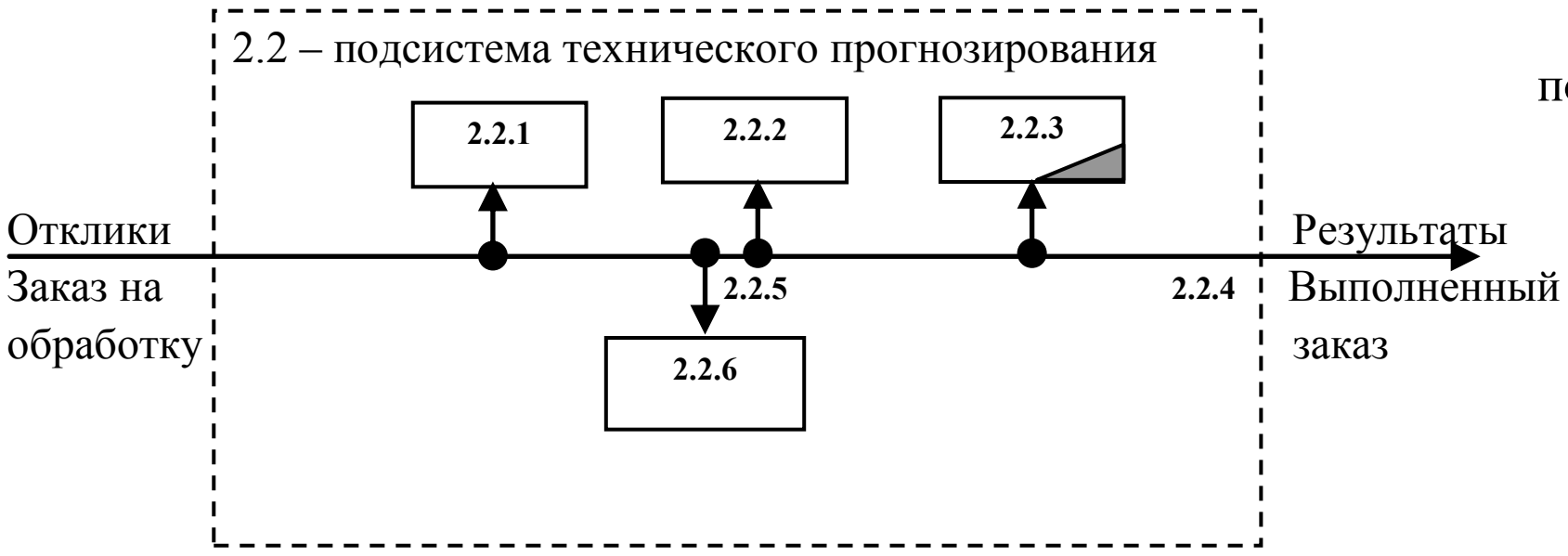

*Второй прототип 2 ранга* – подсистема технического прогнозирования. Блоки:

2.2.1 – формирования выборки, 2.2.2 – моделирования невязки, 2.2.3 – определения детерминированной основы процесса с использованием распознавания образов, 2.2.4, 2.2.5 – интерфейсы, 2.2.6 – прогнозирования и определения вероятностных параметров.

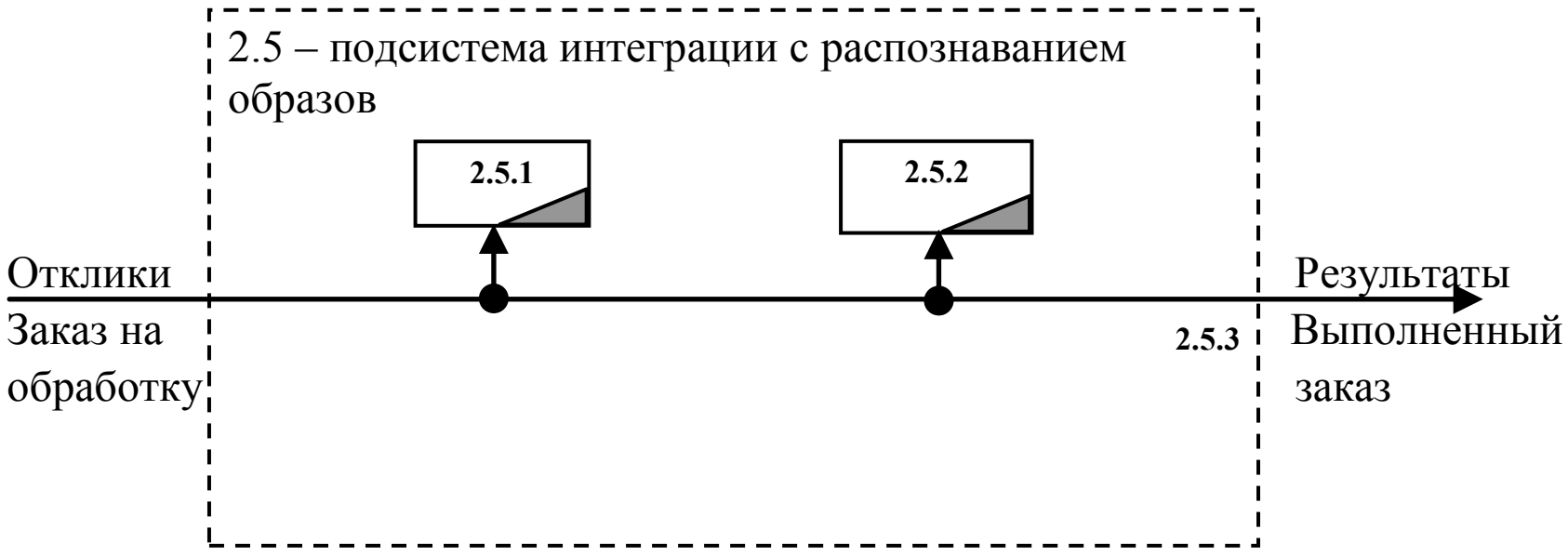

*Третий прототип 2 ранга* – подсистема интеграции с распознаванием образов.

Блоки: 2.5.1 – формирования признакового пространства, 2.5.2 - минимизации описаний, 2.5.3 – интерфейс.

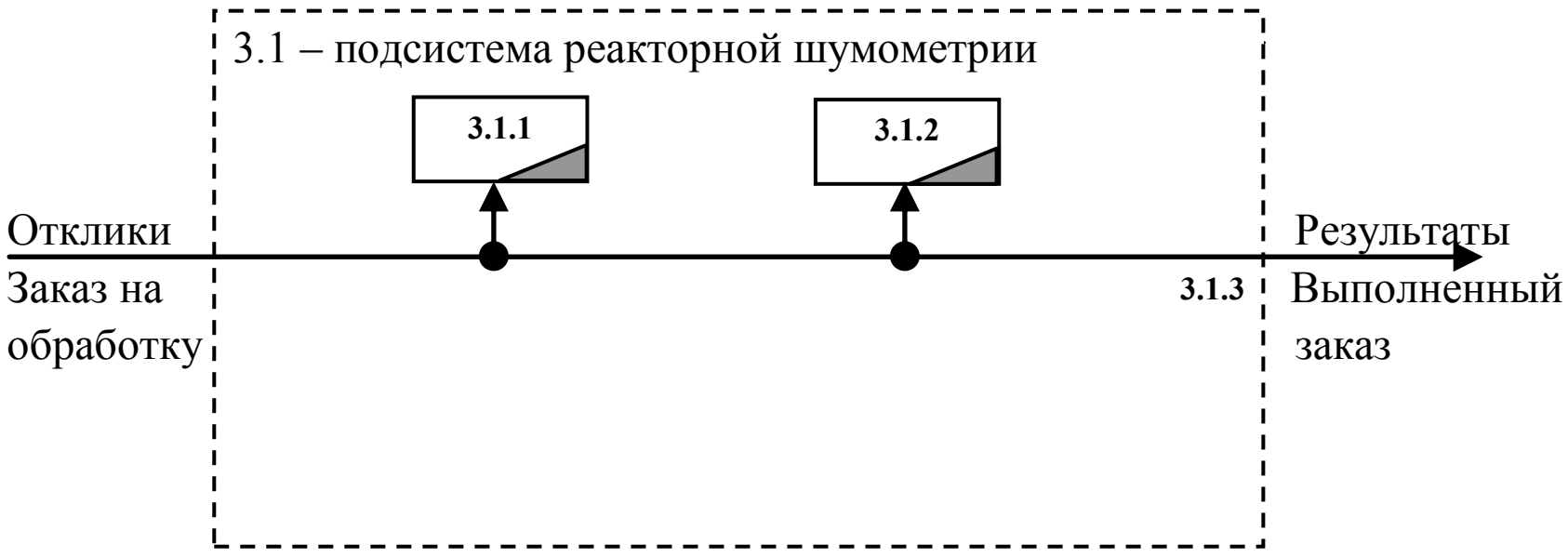

*Четвертый прототип 2 ранга* – подсистема реакторной шумометрии. Блоки:

3.1.1 – формирования описаний процесса, 3.1.2 – диагностики состояния процессов и оборудования, 3.1.3 – интерфейс.

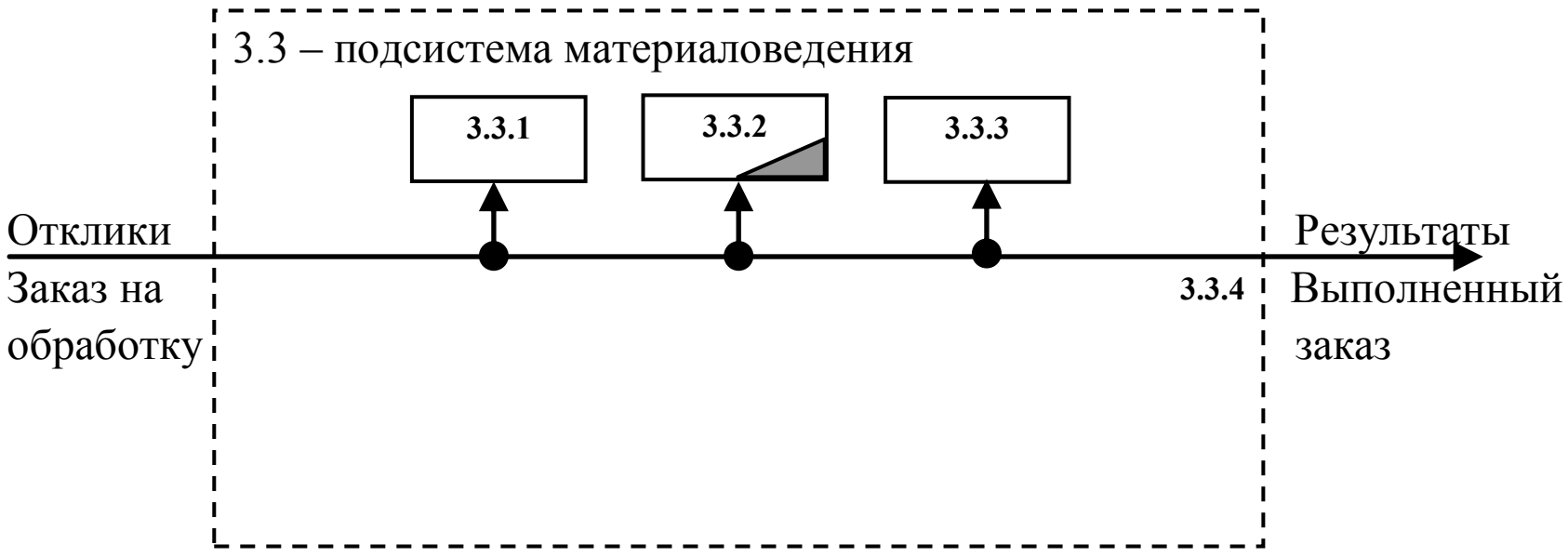

*Пятый прототип 2 ранга* – подсистема материаловедения. Блоки:

3.3.1 – вычисления невязки, 3.3.2 – моделирования невязки с использованием распознавания образов, 3.3.3 – нахождения детерминированной основы и прогнозирования, 3.3.4 – интерфейс.

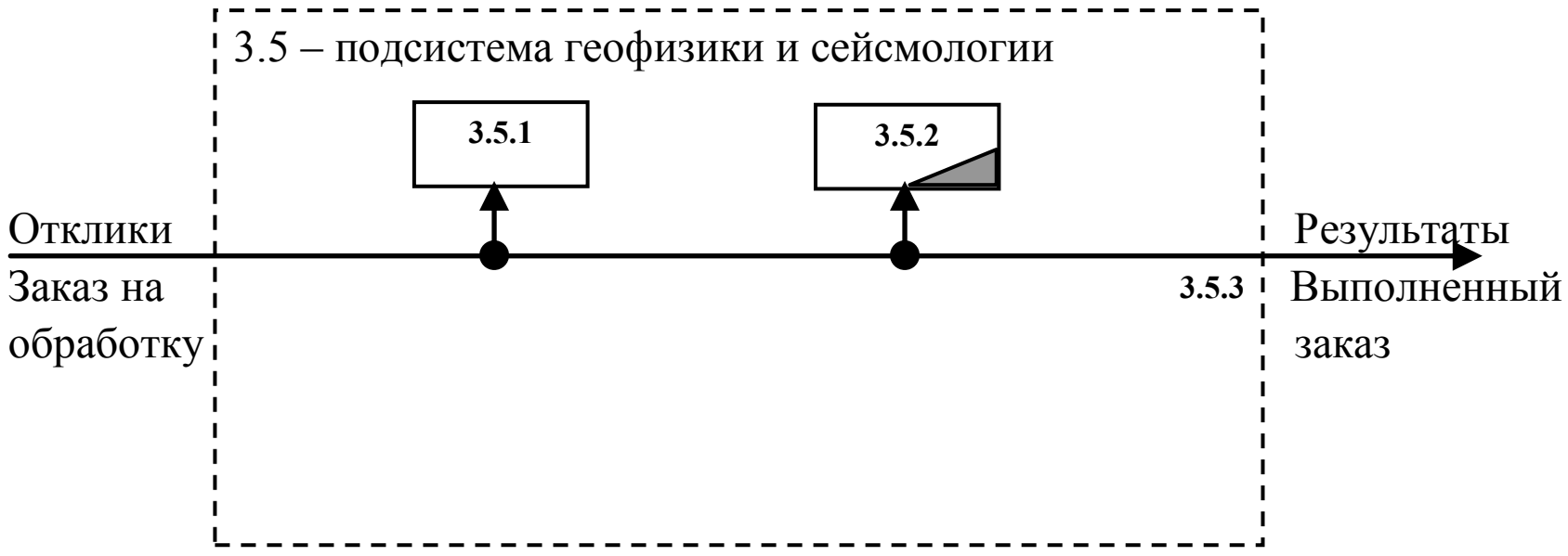

*Шестой прототип 2 ранга* – подсистема геофизики и сейсмологии. Блоки:

3.5.1 – формирования описаний, 3.5.2 - моделирования и прогнозирования, 3.5.3 – интерфейс.

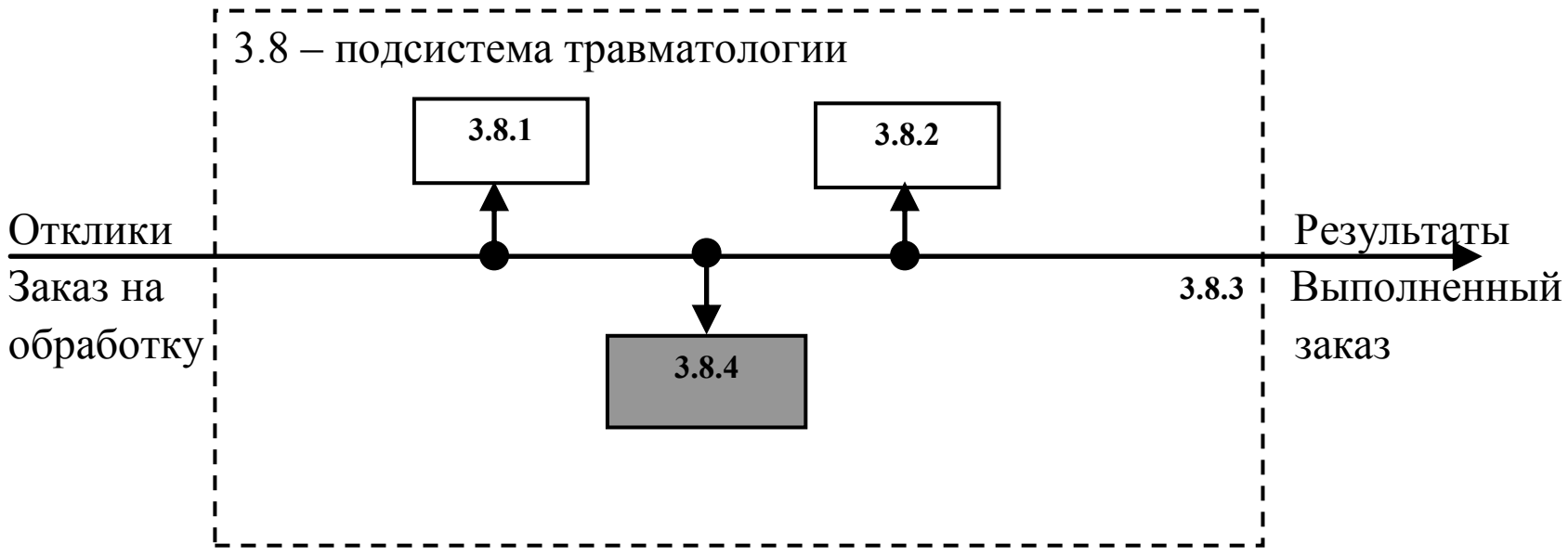

*Седьмой прототип 2 ранга* – подсистема травматологии. Блоки:

3.8.1 – формирования описаний, 3.8.2 – вычисления статистик, 3.8.3 – интерфейс, 3.8.4 – комплексного спектрального исследования.

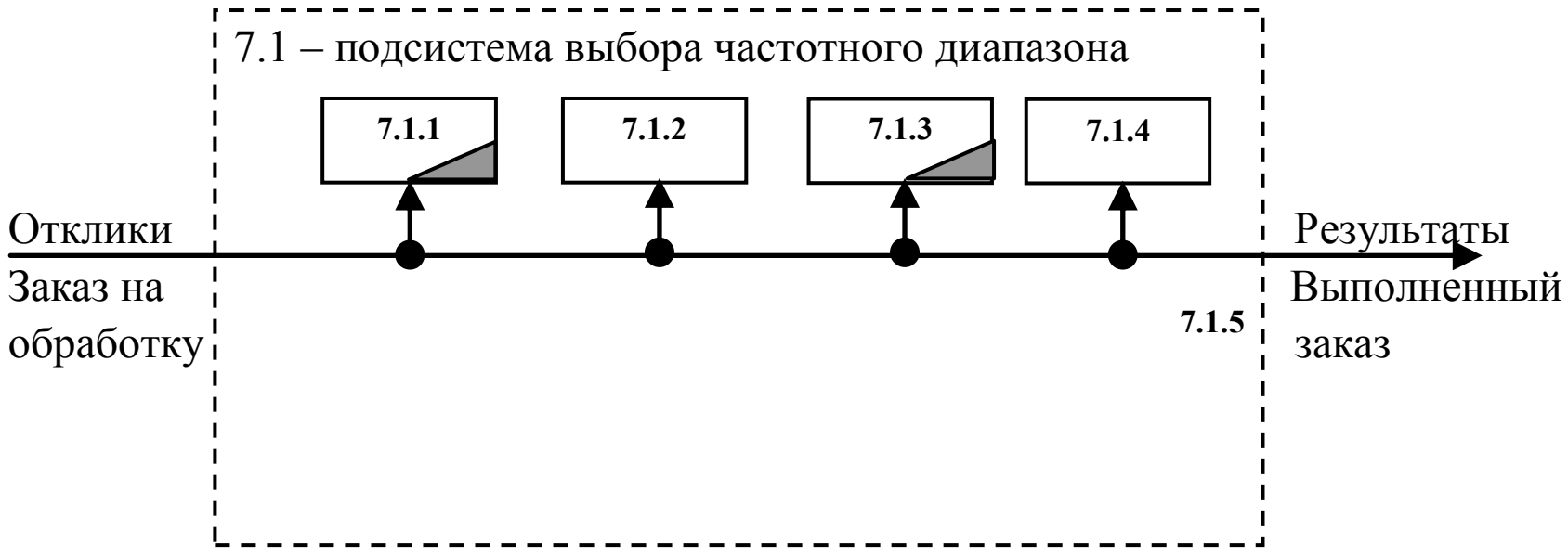

*Восьмой прототип 2 ранга* – подсистема выбора частотного диапазона. Блоки:

7.1.1 – моделирования откликов повышением разрешения в СНЧ диапазоне, 7.1.2 - формирования признаков, 7.1.3 – минимизации описаний, 7.1.4 – принятия решений с использованием распознавания образов, 7.1.5 – интерфейс.

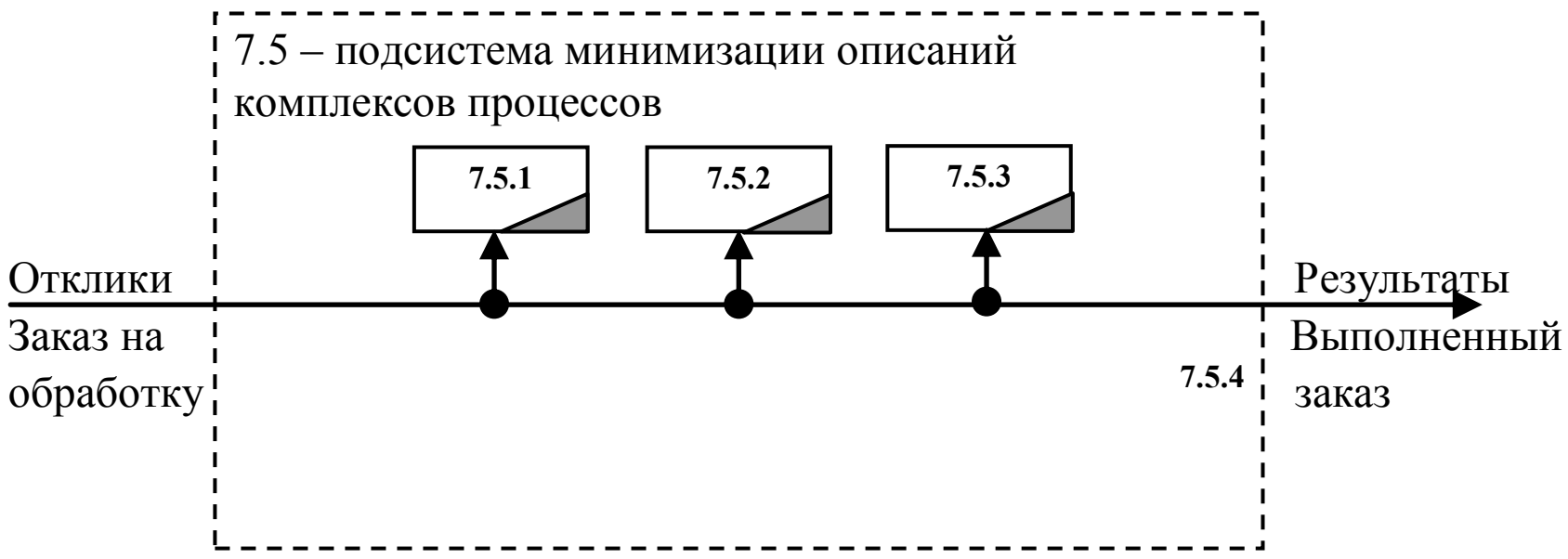

*Девятый прототип 2 ранга* – подсистема минимизации описаний комплексов процессов.

Блоки:

7.5.1 – формирования выборок омплексов процессов из генеральной совокупности, 7.5.2 – формирования признакового пространства, 7.5.3 – минимизации описаний и выбора комплексов процессов, 7.5.4 - интерфейс.

Рис.4. Схема прототипов и предлагаемых решений второго ранга (**подпроект 2.1.2**).

# Проект 2.2 Алгоритмическое моделирование КОС, ее систем и подсистем

В составе проекта 2.2 выполнено 2 подпроекта, представленных ниже на рис.5 и 6.

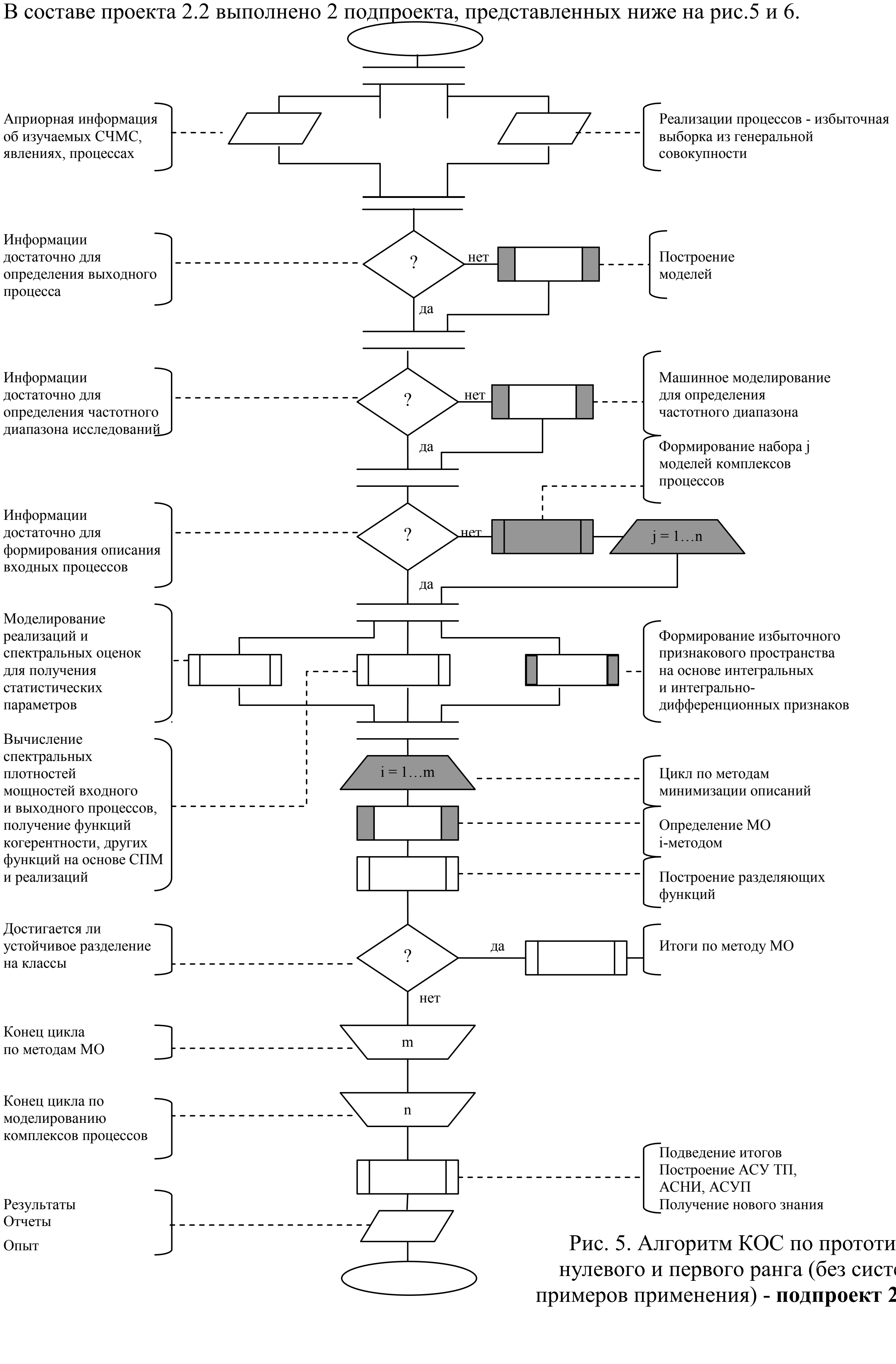

Рис. 5. Алгоритм КОС по прототипам нулевого и первого ранга (без системы примеров применения) - **подпроект 2.2.1**

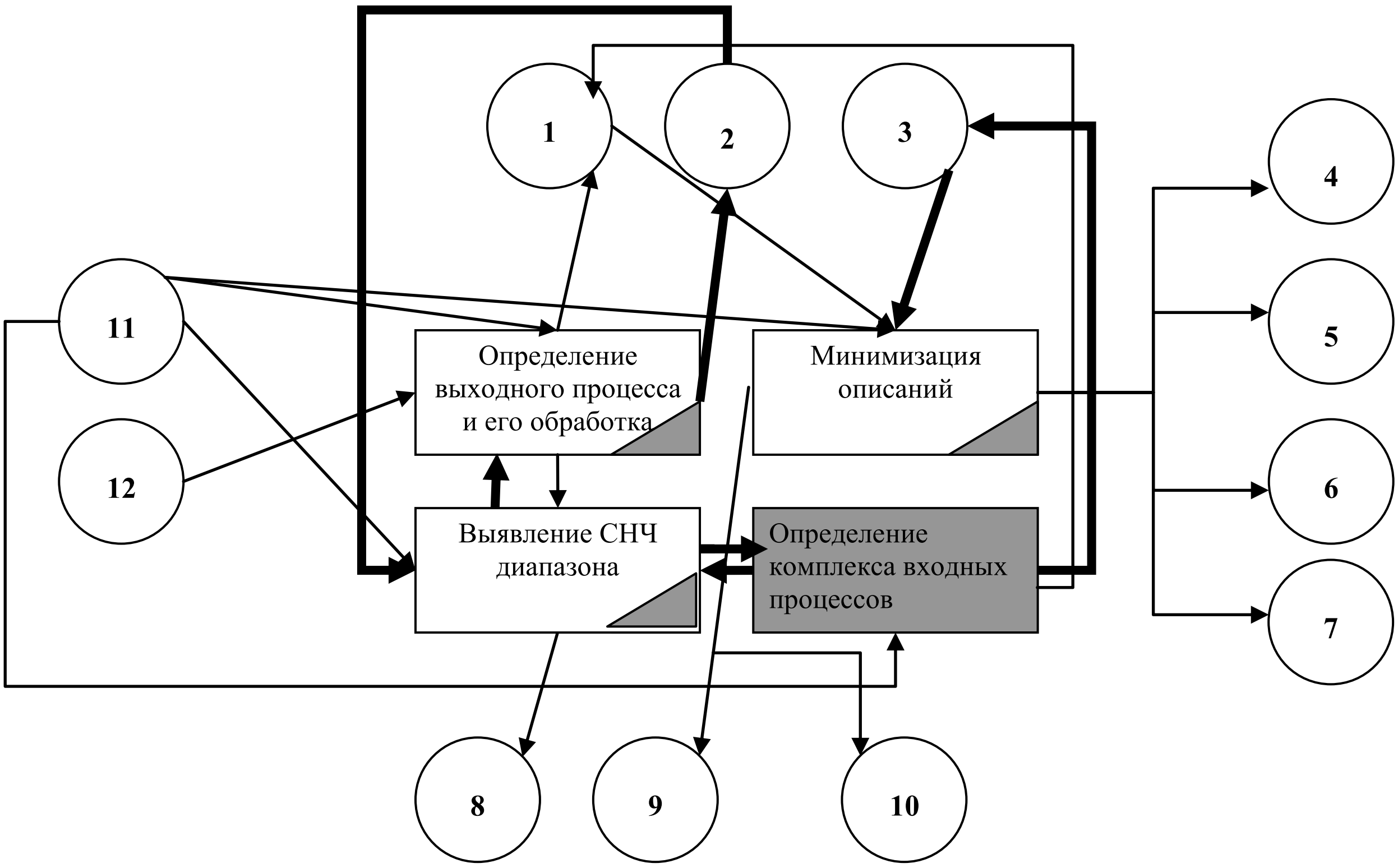

Рис. 6. Схема функционирования КОС (**подпроект 2.2.2**):

1 – спектр и его функции, признаки; 2 – модели СЧМС, явлений или процессов; 3 – модели комплексов процессов; 4 – АСУ ТП; 5 – АСНИ; 6 – АСУП; 7 – новое знание; 8 – неинформативные частотные диапазоны; 9 – отбракованные признаки; 10 – неинформативные процессы; 11 – априорная информация; 12 – реализации процессов.

## Проект 2.3 Информационная модель КОС и программное обеспечение

В составе проекта 2.3 выполнено 2 подпроекта.

Модели КОС в частных случаях основываются на представлении о взаимозависимости процессов и когерентности технологических параметров.

В общем виде суперпозиция процессов различной физической природы может быть сформулирована следующим образом:

$$\begin{cases} a(\tau) = \alpha b(\tau) + \beta c(\tau) + \ldots + \lambda z(\tau) + d(\tau) \\ d(\tau) \le \alpha b(\tau) + \beta c(\tau) + \ldots + \lambda z(\tau) \end{cases} \qquad (1)$$

где *a(τ)* - регистрируемый процесс на выходе СЧМС, *b(τ)* - основной (в смысле выбранной частной модели) процесс на входе, *c(τ)*,…, *z(τ)* - второстепенные процессы на входе, *d(τ)* - моделируемый процесс для учета стохастической природы процесса, α, β,…, λ - функции, учитывающие динамические связи между процессами, в линейном случае - коэффициенты пропорциональности.

Основная идея информационной модели (1) заключается в том, что регистрируя процессы *b(τ)* и *c(τ),…, z(τ),* можно с помощью ФК получить количественную оценку взаимозависимости процессов *a(τ)* и *b(τ)*, *c(τ),…, z(τ)* и тем самым – информацию о состоянии сложной технической системы в целом (**подпроект 2.3.1).**

Разработанный пакет программного обеспечения для КОС (**подпроект 2.3.2)** составляет 38 программ на фортране, оформленных процедурами SUBROUTINE и условно объединенных в следующие блоки:

- программы тестирования исходных реализаций, включающие тесты эргодичности и стационарности, многоцелевую программу МВВКП с подпрограммой решения системы линейных уравнений методом Гаусса, удаление линейного и нелинейного тренда, сортировка методом Шелла, нормирование реализаций к единичному среднеквадратичному отклонению и центрирование относительно среднего;
- программы получения СПМ и моделирования, состоящие из программ БПФ и одновременного БПФ двух реализаций, корректировки на доминирующую частоту, получения биспектров и кепстров для анализа нестационарных процессов, СПМ, взаимной, квадратурной и синфазной СПМ и ее модулей, ступенчатого сглаживания по частоте или ансамблю, улучшения спектра с помощью методов решения некорректно-поставленных задач А.Н.Тихонова [32)] и цифровой фильтрации;
- программы нахождения функций реализаций и СПМ: корреляционные и взаимные корреляционные функции, амплитудно- и фазо-частотные функции, функции обычной, частной и множественной когерентности;
- программы распознавания образов: МО и построение разделяющих функций методом потенциальных функций.

При разработке программ, по мере возможности, соблюдалось требование уменьшения объема занимаемой оперативной памяти путем сокращения количества массивов и переменных. Уменьшение времени вычислений осуществлено использованием быстродействующих и более прогрессивных методов и алгоритмов вычислительной математики, а также сокращением и объединением циклов, выносом за циклы вычислений констант, использованием стандартных функций фортрана. Программы снабжены сервисными модулями для управляемой и самостоятельной печати графических, текстовых и цифровых данных и комментарием для удобства чтения и использования.

Использование пакета ПО заключается в подборе программ и образовании целевых комплексов – систем шумометрии, реализующих тот или иной диагностический метод. Предусмотрена смысловая защита от неправильного использования пакета, заключающаяся в определенной последовательности подключения программ в системе. Системы оформляются в виде основных программ MAIN, или новых процедур SUBROUTINE, что более удобно, так как позволяет расширять пакет за счет таких нестандартных модулей.

## ПРОГРАММА 3 – МЕТОДЫ КОМПЛЕКСНОЙ ОБРАБОТКИ СИГНАЛОВ РАЗЛИЧНОЙ ФИЗИЧЕСКОЙ ПРИРОДЫ

Программа включает 3 проекта и 6 подпроектов.

**Проект 3.1 Техническая диагностика**

В составе проекта 3.1 выполнено 3 подпроекта.

**Подпроект 3.1.1** Выбор диагностических процессов

Формализация выбора диагностических процессов и их комплексов основывается на начальном формировании заведомо избыточной выборки из ГС процессов, с последующим нахождением информативных комплексов процессов методами распознавания образов - минимизацией описаний и построением решающих границ.

**Подпроект 3.1.2** Формирование и математическая обработка сигналов

Формирование реализаций процессов и моделирование сигналов проводилось, как для определения частотного диапазона, так и с целью корректной математической обработки сигналов. Разработаны общие принципы формирования первичных данных в СНЧ диапазоне для разных типов процессов.

Реализации всегда имеют конечную длительность, и спектр усеченного на концах сигнала отличается от спектра бесконечной реализации, вследствие перетекания энергии спектральных составляющих и появления ложных деталей, достигающих значительных величин. Этот эффект во временной области уменьшали с помощью наиболее распространенной косинусоидальной функции, рекомендованной в классических трудах Дж. Бендата и А. Пирсола, в частности [31)], искусственно приближающей к нулю концы реализации, и соответствующим корректирующим множителем в спектральном представлении процесса.

При формировании цифровых реализаций путем обработки аналоговых отображений, например, кривых самописцев в **проекте 5.2**, учитывалось, что гибкость к выбору полосы пропускания при выбранном СНЧ диапазона не является преимуществом, в то время, как помехозащищенность такого способа получения реализаций, существенно ниже, по сравнению с цифровой регистрацией сигнала.

Представление информации, накапливаемой в базах данных, в виде временной развертки телеграфной волны «есть событие» - «нет события», также, с определенными особенностями дальнейшей обработки процесса, дает реализацию процессов СНЧ диапазона, что в дальнейшем рассмотрено в **проекте 5.4.**

Наиболее перспективными для КОС в СНЧ диапазоне процессов вида, рассмотренного в **проектах 5.1-5.3** являются методы n-мерного БПФ, а в **проекте 5.4 –** БПУА. Непосредственное использование Фурье-спектров не рассматривается, поскольку цифровая форма реализаций неадекватно отражает непрерывный сигнал, появляются погрешности, связанные с дискретной формой преобразования Фурье.

Предварительная обработка спектрограмм заключалась в определении уровней значимости гипотез стационарности и эргодичности реализаций. В силу принципа неопределенности, уменьшение статистической погрешности увеличивает ошибку смещения и особенности спектра становятся менее различимы. Поэтому, необходимость выполнения того или иного требования определялась в каждом случае отдельно с учетом цели исследования. В эксплуатационных задачах, таких, как оперативная диагностика энергофизического состояния теплоносителя (**проект 5.2**), вопросы смещения не являются существенными по сравнению с достоверностью анализа, а в физических исследованиях (**проект 5.1**), как правило, определяющим оказывается требование точности проявления отдельных составляющих спектра. Окончательный выбор степени сглаживания в задачах развитой КОС определялся машинным моделированием, с формализованным принятием решений методами РО. Аналогичное машинное моделирование проводилось с целью нахождения верхней граничной частоты, разрешения и погрешностей обработки.

**Подпроект 3.1.3** Вычисление признаков

Предлагаемых в литературе признаков оказывается недостаточно для решения сложных задач, которым посвящены исследования в **проектах 5.2** и **5.3**, поэтому в **подпроекте 3.1.3** разработан ряд дополнительных, ранее не рассматривавшихся в качестве признаков задачи распознавания, на основе деления признаков на дифференциальные (например, значения СПМ на отдельных частотах), интегральные и интегро-дифференциальные (подробнее рассмотрено в **подпроекте 3.3.1**).

**Проект 3.2 Техническое прогнозирование**

Техническое прогнозирование осуществляется тем надежнее, чем более точно известна детерминированная основа по результатам проведенных исследований. В данном проекте разработана методика применения КОС для распознавания детерминированной основы.

Основываясь на идее информационной модели (1), можно предложить следующий частный случай методики КОС для физических измерений:

- в результате проведения дискретных замеров тех или иных физических величин всегда возникает необходимость аппроксимации экспериментальных данных пробными функциями - формульными непрерывными зависимостями; такая аппроксимация проводится с целью получения коэффициентов теоретических зависимостей или уточнения самих формул, что имеет самостоятельное значение для исследований; аппроксимация проводится различными рекуррентными процедурами, наиболее распространенной является МНК, основанный на гипотезе о нормальном распределении ошибок эксперимента, как модельных, так и аппаратных; поскольку нормальное распределение встречается на практике значительно реже, чем лапласовское [29], то более универсальным методом аппроксимации является МВВКП;

- после аппроксимации формируется невязка, как разница между измеряемым процессом и его теоретическим представлением, являющаяся в нашем случае также исследуемым выходным процессом информационной модели (1) с искажениями, внесенными несовершенством модели и метода аппроксимации;

- дальнейший подход КОС заключается в получении СПМ невязки и сравнении с модельным “образом”, соответствующим методу обработки данных; полученная таким образом первичная информация позволяет в дальнейшем определить комплекс процессов – например, это могут быть невязка, передаточная функция аппаратуры и случайная, в смысле определенной модели распределения, ошибка измерений; затем проводится формирование признакового пространства, МО, вычисление ФЧК или ФМК и их интерпретация;

- в результате применения методов КОС удается получить информацию не только о погрешностях эксперимента и метода обработки данных, но и о природной, детерминированной основе процессов, ранее не предсказанную теорией, и тем самым развить представления о процессах.

**Проект 3.3 Распознавание образов**

В составе проекта 3.3 выполнено 3 подпроекта.

**Подпроект 3.3.1** Формирование признакового пространства

Функции случайных процессов являются интегральными характеристиками, описывающими процесс в целом, и в этом плане перспективны, как для построения признакового пространства, так и анализа взаимовлияния комплексов разнотипных процессов.

Интегральные описания не зависят от локальных, не значимых изменений входного образа и устойчиво отражают общие, присущие физике процесса особенности, отличающие данный образ явления от любого другого, реализуемого в ГС процессов. Дальнейшим развитием идей формирования интегральных описаний для КОС является использование интегро-дифференциальных описаний и соответствующих признаков – функций когерентности.

В подпроекте изучены вопросы нахождения и применения корреляционных и взаимных корреляционных функций, амплитудно- и фазо-частотных функций, биспектров, функций обычной, частной и множественной когерентности. Сделан вывод о наибольшей перспективности в КОС функций когерентности (ФК), обычно применяемых в оптических приложениях. Математический аппарат этих функций достаточно разработан, сложности возникают лишь на этапе интерпретации результатов, особенно в отношении изучения взаимовлияния четырех и более разнородных процессов с помощью ФМК. ФК отличаются также своей помехозащищенностью, они тесно связаны с передаточной и амплитудно-частотной функциями исследуемого процесса. Значения ФК на отдельных частотах являются интегро-дифференциальными характеристиками, поскольку характеризуют как процесс в целом, так и его особенности.

**Подпроект 3.3.2** Минимизация описаний

В данном подпроекте методы МО и построение разделяющих функций рассмотрены на модельном примере – регистрируемых на разных уровнях мощности флюктуациях давления в исследовательской ЯЭУ. Сформированное с необходимой избыточностью 10-мерное пространство признаков включало статистические моменты различных порядков, вплоть до 4-го (моменты более высоких порядков неустойчивы [29)]): медиана, коэффициент корреляции, дисперсия, асимметрия, эксцесс, частота Райса, семиинварианты 3-го и 4-го порядков, центральный момент 4-го порядка и коэффициент вариации. Признаки являются интегральными и характеризуют входной образ в целом, что дает основу для получения методики формирования описаний с широким диапазоном применимости. В частности, заметим, что рекомендованные для исследования электроэнцефалограмм интегральные признаки «активность», «подвижность» и «сложность» [7] являются аналогами интегрального уровня шума и частоты Райса.

На модельном примере был изучен ряд известных [26] методов минимизации описаний – по разрешающей способности, корреляционным моментам, числу разрешаемых споров, случайному поиску с адаптацией, главным компонентам (метод Карунена-Лоэва), а также ряд более простых критериев, позволяющих обоснованно уменьшать размерность пространства признаков. Установлено, что недостатком большинства перечисленных методов, определяющим их многообразие, является узкая область применения и неустойчивость результатов при добавлении или исключении отдельных признаков в более сложных задачах. Различные методы МО обладают также специфическими недостатками, потребовавшими разработки новых методов МО применительно к задачам КОС - метода МО по разрешающей способности с использованием идеи случайного поиска с адаптацией, а также метода МО, дающего наиболее устойчивые результаты при гарантированном разделении классов – метода МО по дисперсии коэффициентов разделяющих функций (подробнее эти методы изложены в **подпроекте 4.1.2**).

## ПРОГРАММА 4 – ИНТЕГРАЦИЯ КОС С РАСПОЗНАВАНИЕМ ОБРАЗОВ

Программа включает 2 проекта и 5 подпроектов. Распознавание образов нашло применение в проектах данной программы следующими основными элементами – методы минимизации описаний, формирования признакового пространства и формализованного принятия решений.

**Проект 4.1 Выбор частотного диапазона**

В составе проекта 4.1 выполнено 2 подпроекта.

**Подпроект 4.1.1** Моделирование сигналов в СНЧ диапазоне

Моделирование сигналов проводилось с целью повышения разрешения в СНЧ диапазоне, поскольку слишком широкий частотный интервал не дает нужного разрешения и детального проявления СНЧ пиков.

Окончательный выбор верхней граничной частоты и, соответственно, разрешения в задачах развитой КОС определялся машинным моделированием, с формализованным принятием решений методами РО.

**Подпроект 4.1.2** Минимизация описаний

Минимизация описаний для выбора частотного диапазона в данном подпроекте проводится разработанными методами, указанными в **подпроекте 3.3.2.**

Метод МО по разрешающей способности с использованием случайного поиска с адаптацией выражается следующей зависимостью:

$$\max L\left(X^{i,j}, X^{k}\right) - L\left(X^{i,j}, X^{j}\right) < \boldsymbol{\Delta}, \qquad (2)$$

где $X^{j}$ – эталонное описание j-го класса; $X^{ij}$ – i-тое описание j-го класса;

$L(X^{ij}, X^{j})$ – мера сходства описаний (например, статистика $D^2$);

Δ – пороговое значение разрешающей способности (определяется из расчета возможной статистической ошибки в нахождении признаков).

Метод состоит из следующих этапов:

1) нахождение $L(X^{ij}, X^{k})$ и $L(X^{ij}, X^{j})$ при последовательном исключении каждого признака;

2) сравнение результатов. Например, исключение некоторого признака привело к более заметному улучшению разделения, чем исключение любого другого признака. В этом случае признак дальше не рассматривается, и первый этап повторяется с оставшейся совокупностью признаков;

3) минимизация продолжается до тех пор, пока для всех классов не выполнится условие (2), или дальнейшая минимизация не приведет к ухудшению разделения. Тогда производится добавление последнего исключенного признака и одного из исключенных ранее, выбираемого случайным образом. Последняя процедура вызвана тем, что различные результаты дает не только исключение отдельных признаков в разной последовательности, но и последовательность исключения любой их пары.

Метод МО по дисперсии коэффициентов разделяющих функций, дающий наиболее устойчивые результаты при гарантированном разделении классов, основан на том, что большая дисперсия коэффициента признака у разделяющих функций, построенных для нескольких классов образов, означает значительное изменение признака при переходе от класса к классу, т.е. он является информативным.

Таким образом, потенциальная функция вида:

$$K(X, X_k) = 1 + 4P_1 x_{k1} + 4P_2 x_{k2} + ... + 4P_j x_{kj} + ... + 4P_m x_{km} \quad , \qquad (3)$$

где $4x_{kj}$ - искомые коэффициенты разделяющих функций (первый индекс означает класс образов, второй – номер признака), $P_j$ - $m$ признаков пространства описаний, позволяет построить вариационный ряд по дисперсии коэффициентов $x_{kj}$ и получить наиболее информативное описание в смысле данного метода МО.

**Проект 4.2 Минимизация описаний комплексов процессов**

В составе проекта 4.2 выполнено 3 подпроекта.

**Подпроект 4.2.1** Формирование выборок комплексов процессов

Количество выборок из ГС наборов комплексов процессов на начальном этапе формируется с максимально возможной избыточностью, ограничиваемой только возможностями измерительной аппаратуры. Априорные представления в данном случае не должны накладывать ограничений, в противном случае последующее моделирование методами РО оказывается неполным и не позволяет выявить взаимовлияние процессов в информационной модели КОС (1).

**Подпроект 4.2.2** Формирование признакового пространства

Признаковое пространство в данном подпроекте формируется на основе интегральных (**подпроект 3.3.2**) и интегро-дифференциальных (ФК) признаков, поскольку требуется получение наиболее информативных характеристик процессов в целом. В самом деле, функция когерентности, полученная по значениям спектральных плотностей на отдельных частотах двух или более образов, характеризует как частотные, локальные изменения входного образа, так и общие тенденции изменения исследуемой группы образов. Использование таких функций позволяет проводить комплексное изучение выходного процесса (1) по ряду входных сигналов различной физической природы.

**Подпроект 4.2.3** Минимизация описаний и выбор комплексов процессов

Минимизация описаний в данном подпроекте проводится методами, подробнее изложенными в **подпроекте 3.3.2.** Выбор комплексов процессов осуществляется с помощью известного в РО алгоритма принятия решений построением решающих границ по методу потенциальных функций (3).

## ПРОГРАММА 5 – ПРИМЕРЫ ПРИМЕНЕНИЯ КОС

Программа состоит из 4 проектов и 9 подпроектов.

### Проект 5.1 Моделирование невязки

В составе проекта 5.1 выполнено 3 подпроекта.

**Подпроект 5.1.1** Пропускание через вещество нейтронного потока

Целевой функцией обработки аппаратурных линий сцинтилляционного спектрометра быстрых нейтронов с кристаллом стильбена в качестве датчика являлось получение функций пропускания. Проведено выяснение количественного влияния вида функции распределения невязки на сглаживание реальных аппаратурных распределений, а также рассмотрение ряда функций на лучшее соответствие детерминированной основе процесса. Результаты внедрены в прикладных исследованиях пропускания нейтронного потока через различные виды реакторных материалов с целью уточнения физических констант.

**Подпроект 5.1.2** Газовыделение из материалов биологической защиты

Теоретические исследования показали, что радиационно-химический выход водорода из полиэтилена, входящего в материал биологической защиты ЯЭУ, определяется несколькими процессами, зависящими от поглощенной дозы. Проблема заключалась в получении алгоритма распознавания границ радиационно-химических процессов и нахождении точных значений кинетических параметров. В данном подпроекте для получения разделяющей функции границ процессов были выбраны комплексы экспериментальных кривых, полученные при облучении в ЯЭУ образцов полиэтилена с 2,5% содержанием $B^{10}$. Моделированием с помощью МВВКП разработана методика распознавания кинетических параметров газовыделения из полиэтилена, позволяющая определить границы процессов по поглощенной дозе, без разрушения образца, непосредственно в ходе эксперимента по регистрации газовыделения. Получено условие разделения экспонент, описывающих различные радиационно-химические процессы, такие как образование поперечных связей в аморфной и кристаллической фазах полиэтилена и транс-виниленовая ненасыщенность. Методика используется для уточнения констант протекания радиационно-химических реакций и прогнозирования газовыделения из полиэтилена с различным содержанием наполнителей.

**Подпроект 5.1.3** Зондовые характеристики низкотемпературной плазмы

Решение уравнения Больцмана, используемое для описания экспериментальных распределений электронов по энергиям однозондовой характеристики, дает общий вид пробных функций, которые могут быть использованы для аппроксимации разностной двухзондовой характеристики, решение для которой найти не удается. Моделирование невязки с помощью МВВКП и формирование признакового пространства с последующим распознаванием детерминированной основы процесса показали, что наилучшим образом экспериментальный материал описывается пробной функцией вида

$$Y = a_1 \exp( a_2 x + a_3 x^2 + a_4 x^3 ), \qquad (4)$$

что дает важную информацию для понимания физической основы процесса.

**Проект 5.2 Комплексная обработка сигналов ТП ЯЭУ**

Частный случай модели (1) для ТП ЯЭУ можно получить, если считать колебания мощности ЯЭУ $n(\tau)$ результирующим процессом на выходе СЧМС,

а пульсации температуры $t$, давления $p$ и плотности $\rho$ - процессами на входе. Тогда можно записать аналогичную модели (1) математическую модель когерентности теплофизических процессов ЯЭУ:

$$\begin{cases} \delta K_c(\tau) = \int\limits_v \begin{Bmatrix} \left[\alpha_t(t) + \alpha_n(t) + \alpha_\rho(t)\right] \times \left[\delta t(\tau)\psi_t(\bar{r},\tau) + t_0 \delta\psi_t(\bar{r},\tau)\right] + \\ + \alpha_p(p)\left[\delta p(\tau)\psi_p(\bar{r},\tau) + p_0 \delta\psi_p(\bar{r},\tau)\right] + \\ + \alpha_\rho(\rho)\left[\delta\rho(\tau)\psi_\rho(\bar{r},\tau) + \rho_0 \delta\psi_\rho(\bar{r},\tau)\right] \end{Bmatrix} dv + \delta F(\sigma;\upsilon;\chi;\psi_h;\phi;\tau) \\ \\ \delta F(\sigma;\upsilon;\chi;\psi_h;\phi;\tau) \le \delta K_T(\tau) \\ n(\tau) = H(n_0;P) \times K_c(\tau) = H(n_0;P) \times \delta K_c(\tau) \end{cases} \qquad (5)$$

где $\delta$ - означает флюктуационную составляющую соответствующей величины, $K_c(\tau)$ - общая реактивность системы, $K_T(\tau)$ - реактивность, обусловленная действием эффектов реактивности, $\psi(r,\tau)$ - форм-функции полей соответствующих теплофизических величин, $\alpha$ - коэффициенты реактивности, индекс «0» означает стационарное состояние системы, $F$ - составляющая реактивности, обусловленная стохастической природой нейтронно-физических процессов, функция сечений взаимодействия, вероятностей взаимодействия нейтронов с элементами активной зоны, спектра мгновенных и запаздывающих нейтронов, форм-функции нейтронного потока и потока нейтронов; $H(n_0;P)$ - передаточная функция.

При выводе модели когерентности (5) не делалось каких-либо ограничений, учитывающих конструкцию ЯЭУ, что позволяет считать модель достаточно общей и пригодной для разных типов аппаратов. Конкретизация модели осуществляется определением основного процесса на входе и выбором типа ФК. Так, в реакторе с водой под давлением, где появление парогазовой фазы носит эпизодический характер, основным процессом могут служить флюктуации давления $\rho$, а в кипящем аппарате определяющую роль играют флюктуации плотности $p$. В гипотетическом случае важности всех входных процессов или для исследования зависимости выходного и измеряемых второстепенных процессов, в анализе могут быть применены ФМК.

Проверка модели (5) проводилась для частного случая погружного водо-водяного исследовательского реактора с опускным движением воды, пластинчатыми твэлами и узкими каналами тепловыделяющих сборок. Соответствующее заданным условиям упрощение (здесь, для краткости, не приводится) модели (5) приводит к формированию модели КОС с регистрируемыми флюктуациями мощности на выходе системы, основным процессом на входе – пульсациями давления и второстепенным – флюктуациями температуры теплоносителя.

ФЧК флюктуаций давления *p* и мощности *n* при МНК-исключении влияния на основной входной и результирующий процессы шумов температуры *t*, определится выражением:

$$\hat{\gamma}^2_{pn.t} = \frac{\hat{G}^2_{pn}\left[1-\frac{\hat{G}_{pt}\times\hat{G}_{nt}}{\hat{G}_t\times\hat{G}_{pn}}\right]^2}{\hat{G}_p\left[1-\frac{\hat{G}^2_{pt}}{\hat{G}_p\times\hat{G}_t}\right]\times\hat{G}_n\left[1-\frac{\hat{G}^2_{nt}}{\hat{G}_n\times\hat{G}_t}\right]} \qquad (6)$$

где $\hat{G}$ - оценка СПМ соответствующего процесса.

Предсказанное моделью, поведение ФЧК при изменении режимов теплофизического состояния реакторной установки, получило экспериментальное подтверждение, как показано на рис.7.

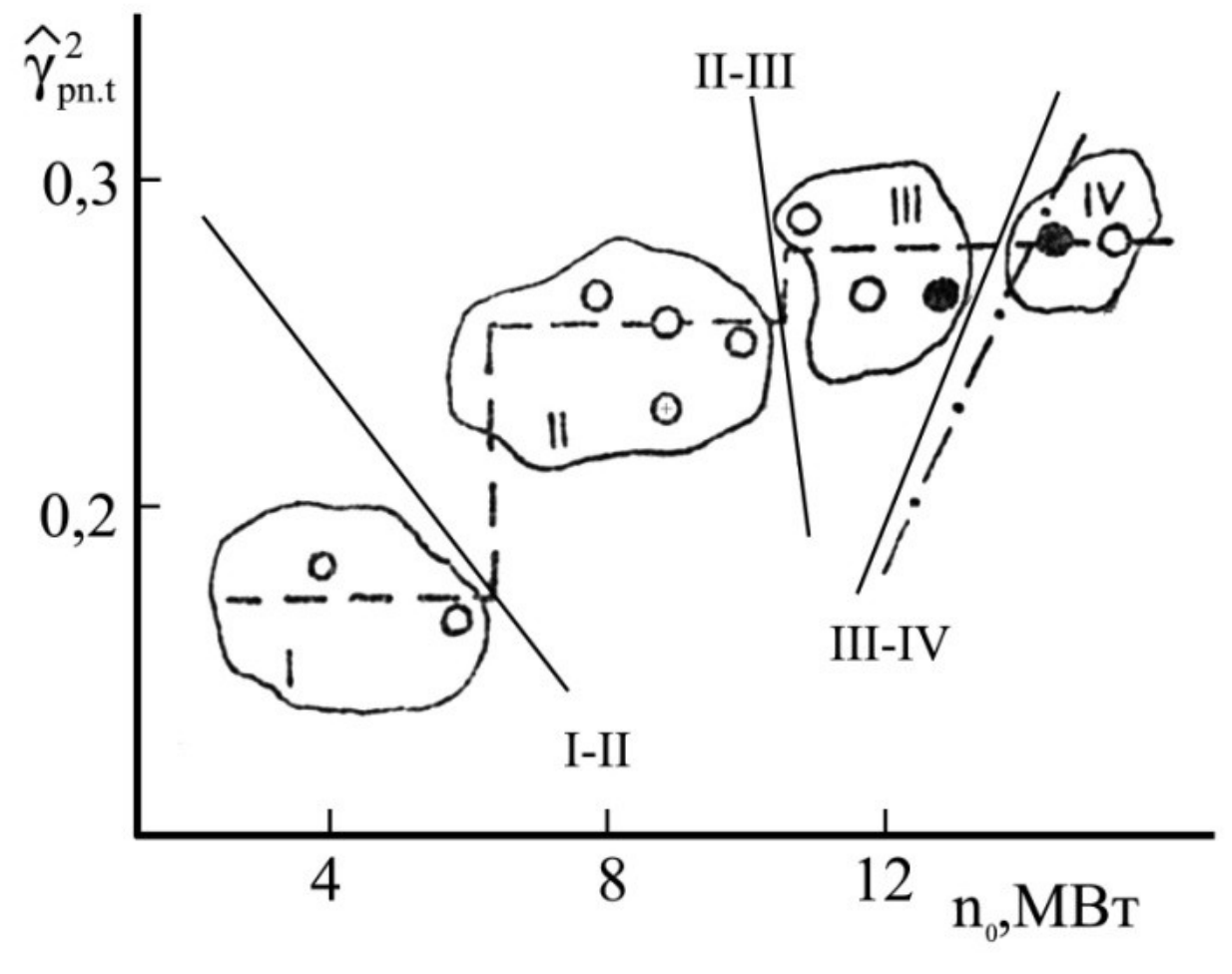

Рис.7. Изменения ФЧК в зависимости от мощностных режимов ЯЭУ
I, II, III, IV – классы теплофизических режимов ЯЭУ, ⊕ - нестационарный режим, ● - неравномерное энерговыделение, сплошной линией показаны разделяющие границы, штрих-пунктиром – разделяющая граница, когда класс IV задан одной точкой на мощности 15,7 МВт, ступенчатым пунктиром показано поведение ФЧК, предсказываемое упрощением модели (5) для конкретного аппарата. СПМ ФЧК получены с $f_B$ = 5 Гц и $\Delta f$ = 0,04 Гц.

## Проект 5.3 КОС оборудования АЭС

В составе проекта 5.3 выполнено 2 подпроекта.

**Подпроект 5.3.1** Флюктуации ТП в технологических трубопроводах

Общий вид модели КОС (1) был апробирован также в задаче определения мощности энергоблока с реактором БН-600 по показаниям расходомеров парогенераторных секций III контура. Предварительный анализ однотипных процессов - СНЧ колебаний расхода, регистрируемого расходомерами в диапазоне до 0,02 Гц, позволил выявить полигармонические колебания расхода через секцию. Дальнейшее изучение флюктуаций с помощью ФЧК показало, что процессы когерентны внутри петли и не когерентны между петлями и связаны с геометрией трубопроводов и работой петлевых насосов. Отсюда были получены рекомендации по индивидуальному для каждой петли получению замеров и устранению неточности в определении мощности при используемых дискретных измерениях расхода.

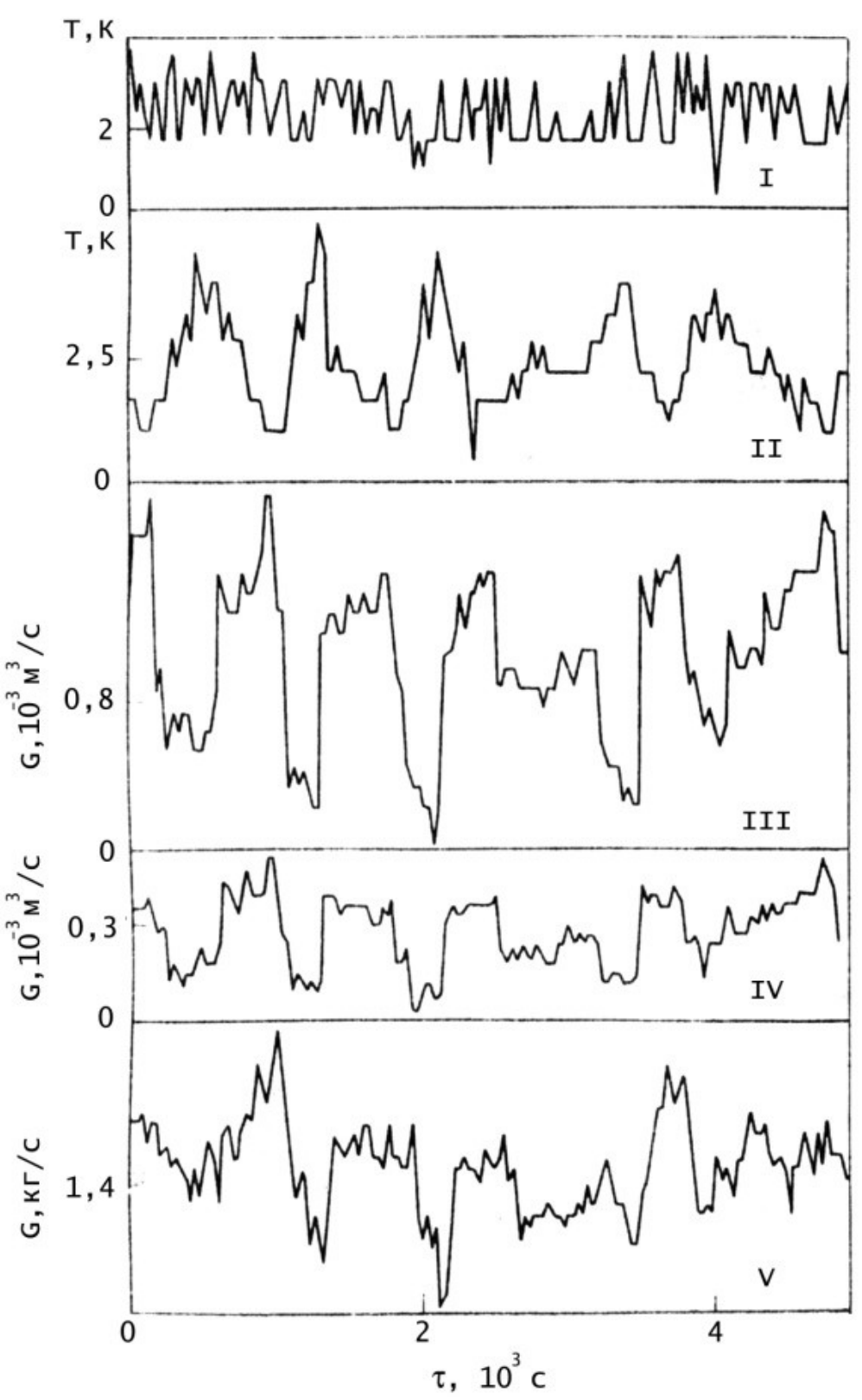

Рис.8. Флюктуации ТП в парогенераторе I – температура натрия на входе в секцию, II – температура натрия на выходе из секции, III – расход питательной воды, IV – секционный расход питательной воды, V – расход острого пара.

В следующей работе данного подпроекта, на основе модели (1) были исследованы комплексы разнотипных процессов – флюктуации давления, температуры и расхода пара, воды и натрия на петлевых парогенераторах ПГН-200М. Реализации процессов, представленные на рис.8, регистрировали штатной аппаратурой с $f_B$ = 0,015 Гц и $\Delta f$ = $1{,}2 \cdot 10^{-4}$ Гц.

Механизм появления СНЧ флюктуаций не может быть связан с аппаратурным эффектом, когда период времени между отсчетами больше периода высокочастотной модуляции потока питательным насосом. В противном случае когерентность флюктуаций температуры натрия и расхода питательной воды (рис.8, II, IV) была бы мала. Сходство структуры реализаций процессов, независимо регистрируемых информационно-измерительными каналами на стороне петли и на входе в секцию парогенератора (рис.8, III, IV), показывает, что реализации отражают общую детерминированную основу процессов, и не являются шумами измерительной аппаратуры.

На основе исследования сделан вывод об определяющем значении флюктуаций расхода питательной воды в формировании общего поля изменений ТП парогенератора. Управление структурными характеристиками флюктуаций расхода питательной воды может использоваться для оптимизации работы парогенераторов в качестве составной части АСУ ТП АЭС.

**Подпроект 5.3.2** Диагностика и прогнозирование состояния конденсационных установок

Аналогичная по своей постановке **проекту 5.1** работа проведена для определения аналитической зависимости потерь мощности $\Delta N$ от снижения вакуума $\Delta P$ при определенных значениях расхода пара $D$ в конденсаторах турбин ВК-100-6 и К-200-130 Белоярской АЭС им.И.В.Курчатова.

Экспериментальные данные функциональной зависимости вида:

$$\Delta N = f(\Delta P, D) \tag{7}$$

аппроксимировали с помощью МВВКП конкурирующими экспоненциальными и полиномиальными функциями. После аппроксимации графического материала, по невязке были найдены критерии качества аналитических моделей – признаки задачи распознавания режима работы системы, и определена искомая аналитическая зависимость.

## Проект 5.4 Адаптация КОС к исследованиям процесса ПТ

В составе проекта 5.4 выполнено 4 подпроекта.

**Подпроект 5.4.1** Алгоритм адаптации КОС

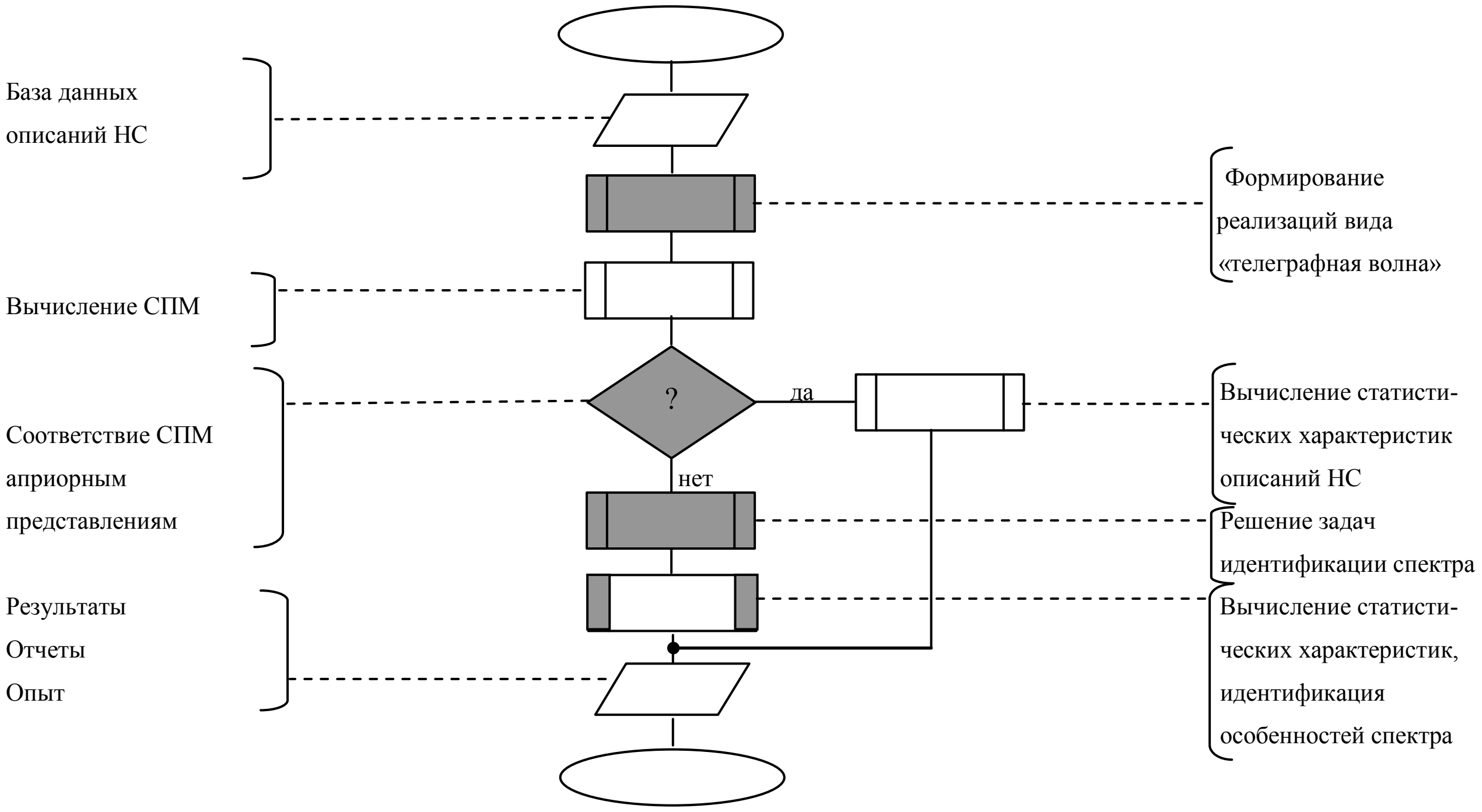

Рис.9. Алгоритм адаптации КОС к исследованиям ППТ

Приведенный на рис.9 алгоритм адаптации КОС к исследованиям процесса производственного травматизма (ППТ) далее подробно рассмотрен в **подпроектах 5.4.2-5.4.4.**

**Подпроект 5.4.2** Обработка сигналов ППТ России и Италии

База данных для исследования ППТ в России была образована на основе анкет НС по прозводственному объединению (11-летний массив данных, содержащий 7000 НС), промышленному предприятию (6-летний массив, содержащий 625 НС) и, в целом, отрасли тяжелого и транспортного машиностроения (1-летний массив отдельных видов НС).

Далее, образованы реализации ППТ, вида, представленного на рис.10 (фрагмент из раздела «производственное объединение»).

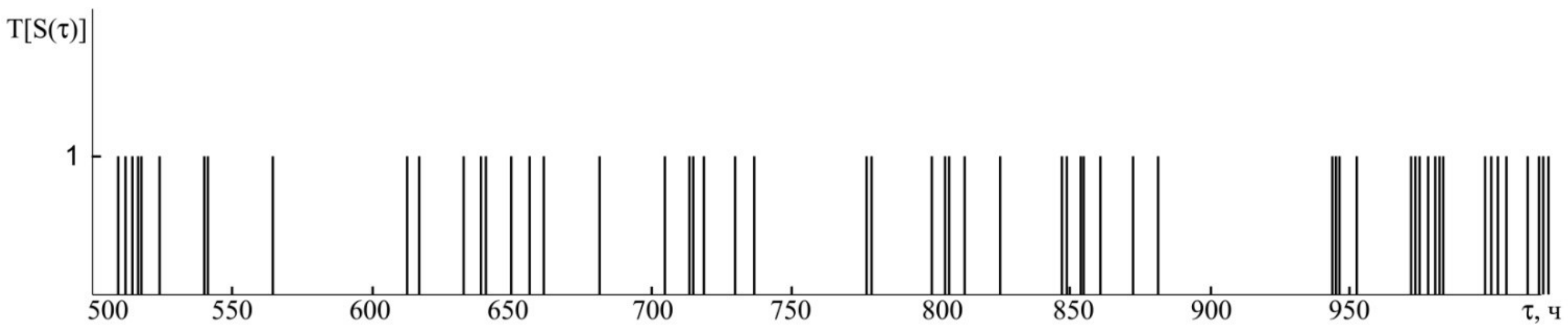

Рис.10. Вид реализаций ППТ базы данных НС по России

Пример полученных в данном подпроекте оценок СПМ в СНЧ диапазоне приведен на рис.11.

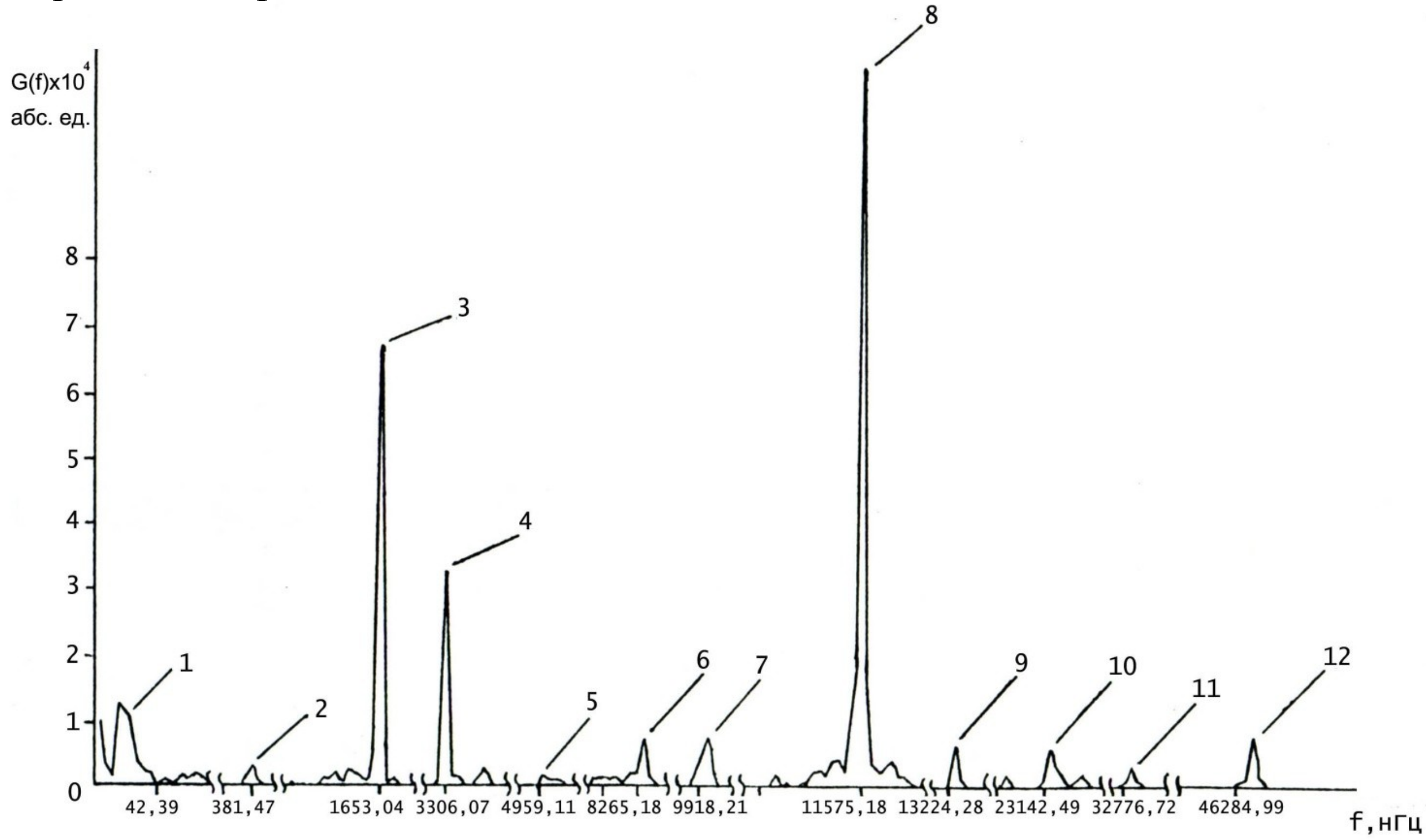

Рис.11. СПМ из раздела базы ППТ России «производственное объединение»

Число точек реализации $N$ =65536, $f_B$ = 138,9 мкГц, $\Delta f$ = 4,238 нГц; периоды гармоник : 1 – 1,24688 лет, 2 – 30,34074 сут., 3 – 7,00171 сут., 4 – 3,50085 сут., 5 – 56,01367 час., 6 – 33,59000 час., 7 – 27,99487 час., 8 – 23,99707 час., 9 – 20,99840 час., 10 – 12,00073 час., 11 – 8,47485 час., 12 – 5,99982 час.

Видно, что линейчатый спектр ППТ не является марковским, более того, не относится к случайным процессам, а является детерминированным полигармоническим процессом.

Для определения статистических характеристик и интерпретации результатов было проведено машинное моделирование годовых и полных разделов базы **подпроекта 5.4.2** для определения основных параметров спектрального анализа: $\Delta\tau$, $\Delta f$, $f_B$, длины реализации. Исследованы вопросы эргодичности и стационарности реализаций $T[s(\tau)]$, уменьшения утечки энергии через боковые лепестки с помощью косинусоидальной функции, влияния дополнения реализации нулями, явления маскировки частот, амплитудной модуляции спектра, а также проведен сравнительный анализ методов БПФ и БПУА.

Установлено, что частотная структура спектров ППТ год от года не меняется при изменении длин реализаций, варьировании сглаживанием, параметрами $\Delta\tau$, $\Delta f$, $f_B$ , методом обработки и при анализе 11-ти различных годовых выборок, даже в тех случаях, когда количество НС в сравниваемых годах отличается ~ в 2 раза. Более того, устойчивость полученных результатов характеризуется тем, что аналогичное исследование базы ППТ «промышленного предприятия», не входящего в «производственное объединение» и расположенного в другом экономическом районе России, показало качественное совпадение спектров ППТ предприятия и объединения. Основные указанные пики видны и на спектрах раздела базы «отрасль в целом».

Варьирование сглаживанием по частоте СПМ (рис.11) показало, что при минимально возможной ошибке смещения $\varepsilon_b$ и максимальном разрешении $\Delta f$, погрешность частотной идентификации особенностей спектра не превышает 0,2% от частоты, что является вполне достаточным и позволяет сделать выбор соответствующих параметров спектрального анализа.

База данных для изучения ППТ Италии была сформирована на основе 1-летней выборки, содержащей 10471 НС, из банка данных ПТ наиболее развитых в промышленном отношении областей Италии – провинций Милана и Бергамо, предоставленных автору I.N.A.I.L.

В базу данных занесены, реализации ППТ, вид которых показан на рис.12.

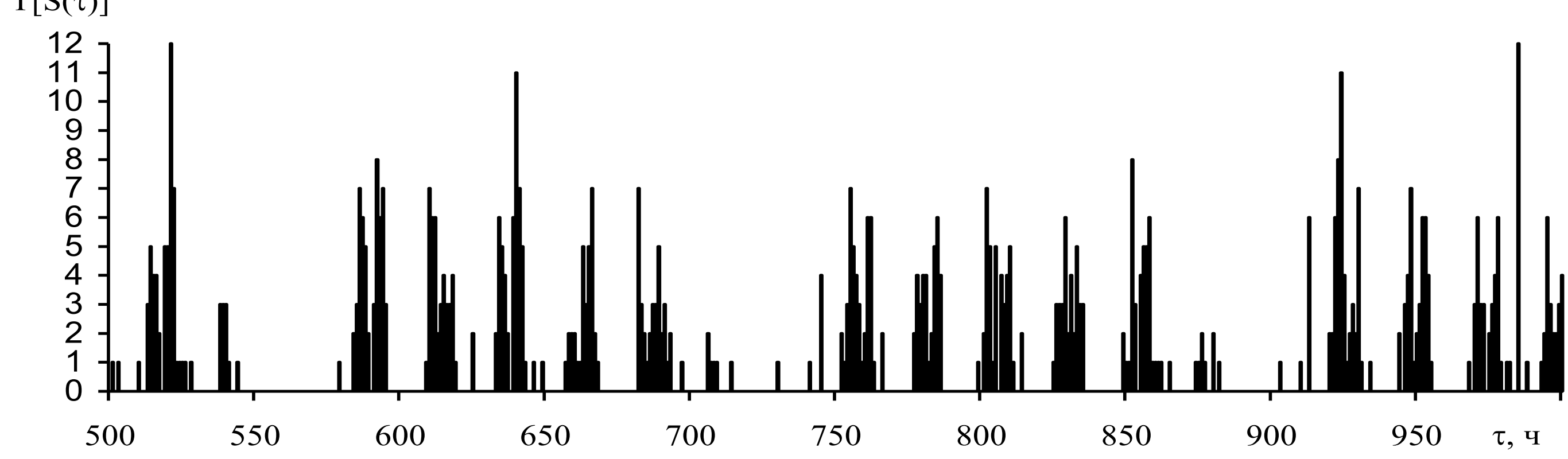

Рис.12. Вид реализаций ППТ базы данных НС по Италии

Насыщенность НС выборки ПТ провинций Бергамо и Милана более, чем в 15 раз выше, чем максимальная из анализируемых по России («производственное объединение», рис.10).

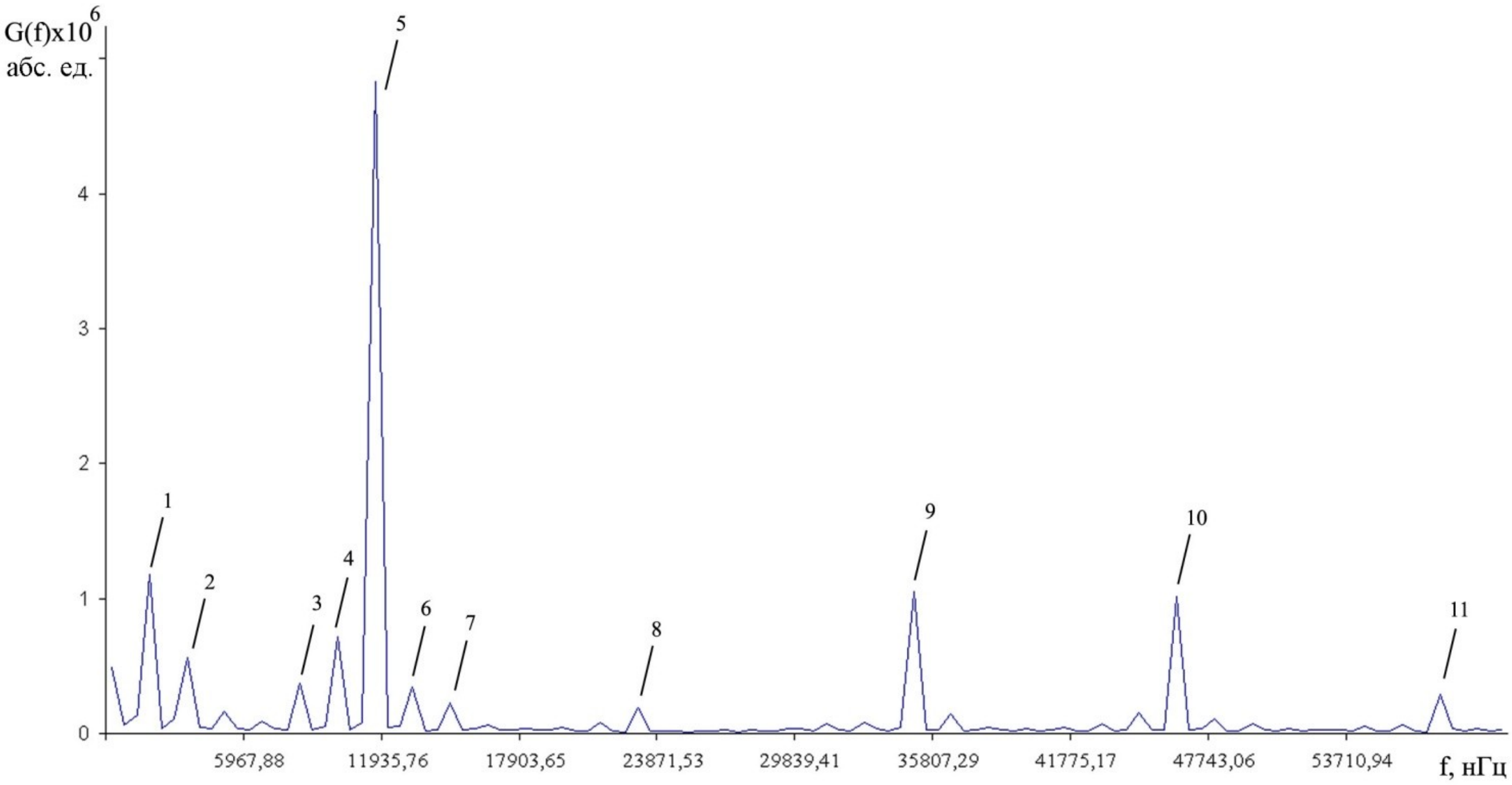

Рис.13. СПМ ППТ Италии

$N = 8192$, $f_B = 138{,}9$ мкГц, $\Delta f = 33{,}9$ нГц; периоды гармоник : 1 – 7,11111 сут., 2 – 3,55555 сут., 3– 34,13333 час., 4 – 28,44 час., 5 – 24,38095 час., 6 – 21,333 час., 7 – 18,96296 час., 8 – 12,19048 час., 9 - 8,0000 час., 10 – 6,02352 час., 11 – 4,83018 час.

**Подпроект 5.4.3** Автоколебательная модель ППТ

В соответствии с моделью (1) представим выходной процесс *T[s(τ)]* СЧМС, как упорядоченную временную последовательность объектов – НС:

$$T[s(\tau)] = S(\tau_1), S(\tau_2),..., S(\tau_j),..., \Delta\tau = \tau_j - \tau_{j-1} \succ 0 \quad (8)$$

где $\tau_1$, $\tau_2$,…, $\tau_j$,… - моменты времени (дискретная временная ось).

Такая последовательность является не только функцией времени, заданной на сетке $\{\tau_i\}_{i=1}^{\infty}$, или, в непрерывном представлении, когда $\Delta\tau \to 0$ – $s(\tau)$, но и имеет смысл функциональной зависимости от времени, обусловленной целым рядом входных процессов – циркадианными, недельными (циркасепдианными), месячными (рис.10,12), инициируемыми не только факторами среды (экзогенными) и эндогенными, т.е. заданными генетически, но и социально-организационными факторами, присущими СЧМС.

Модель ППТ описывает вынужденные перечисленными процессами колебания ППТ вдали от точки естественного минимума, в качестве которого принимаем марковский процесс:

$$G(f) = 2\left|F\{T[s(\tau)]\}\right|^2 = const \qquad (9)$$

где $F$ - знак преобразования Фурье.

Спектр ППТ (рис.11,13) имеет две основные частоты $f_1$=1653,04±2,11 нГц и $f_2$ = 11575,48±2,11 нГц, причем вторая основная частота кратна первой: $f_2=7f_1$. В самом деле, частоты гармоник с номерами 3÷10,12 с высокой степенью точности кратны $f_1$ , а гармоники 8,10,12 – $f_2$. Таким образом, модель ППТ может быть представлена в виде автоколебательной системы, совершающей вынужденные гармонические осцилляции, спектр которых аналогичен набору тонов и обертонов струн, ограниченных с двух сторон с отношением длин 7:1 (рис.14).

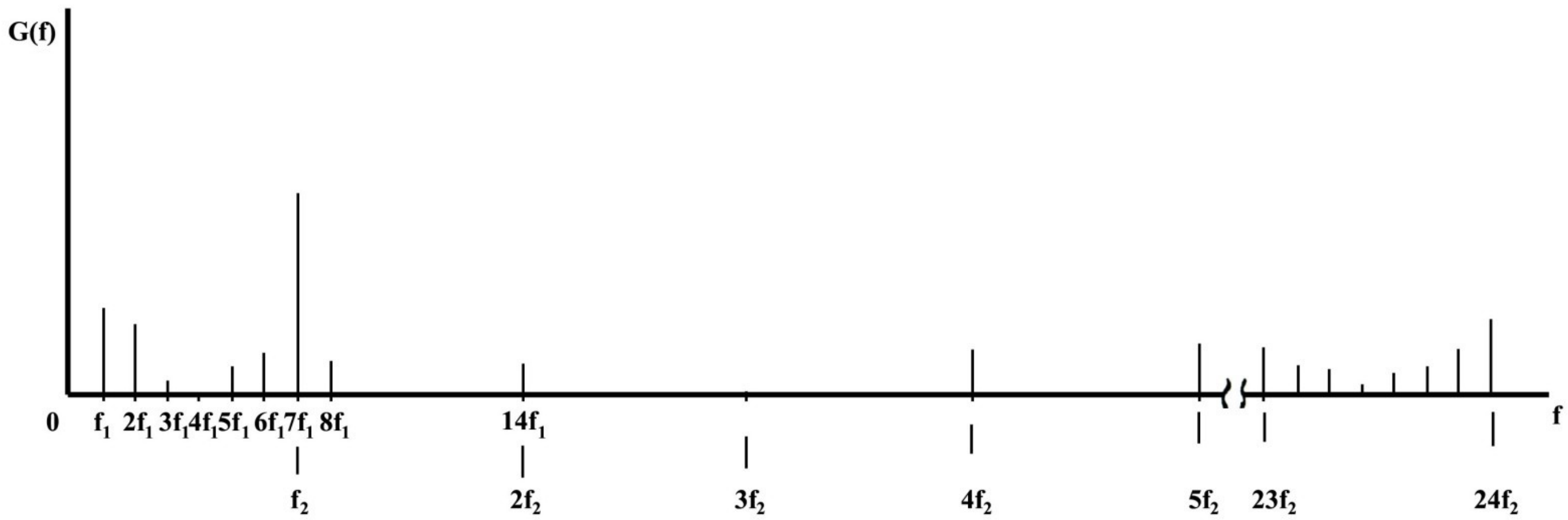

Рис.14. Модель СПМ полигармонического ППТ

Видно, что СПМ содержит три правильные серии: 0÷7$f_1$; 0÷24$f_2$; 23$f_2$÷24$f_2$. Последняя серия, по-видимому, связана со сверткой в низкочастотную область гармоник с периодами осцилляций, меньшими 1 часа, вследствие явления маскировки. Первая серия определяет расщепление основной гармоники $f_1$, имеющей период 7 суток и вторую по величине в спектре амплитуду.

Таким образом, дальнейшие исследования автоколебательной модели в соответствии с моделью КОС (1), должны быть направлены на формирование информативной выборки из генеральной совокупности входных процессов, как исследуемых в хронобиологии, так и обусловленных социальными и организационными периодическими факторами – управляющими воздействиями в СЧМС, с последующим вычислением и интерпретацией корреляционных зависимостей между ППТ и вынуждающими синергетические колебания ППТ хронобиологическими циркадианными факторами, а также соответствующих ФЧК и ФМК.

**Подпроект 5.4.4** Сравнение результатов обработки сигналов ППТ Италии и России

Сравнение статистик НС по базам данных ППТ Италии и России проводилось с целью идентификации циркадианных и недельных ритмов, обнаруженных с помощью применения КОС (рис.15).

Исследование показало, что циркадианные ритмы ППТ определяются психологическими, поведенческими причинами с традиционно рассматриваемым в хронобиологии [33)] набором внешних воздействий и реакций человека. В то же время недельный ритм определяется социально-организационными факторами СЧМС, включая и систему инструктажей, как управляющих воздействий на ПТ.

Данные, представленные на рис.11 и 13, показывают практическое совпадение спектральных характеристик ППТ, полученных в разные годы, на различных по типу и величине предприятиях и отраслях, в разных странах с существенно отличными условиями организации труда. ППТ во всех случаях имеет ярко выраженный детерминированный, полигармонический характер с доминантной циркадианной 24-часовой гармоникой.

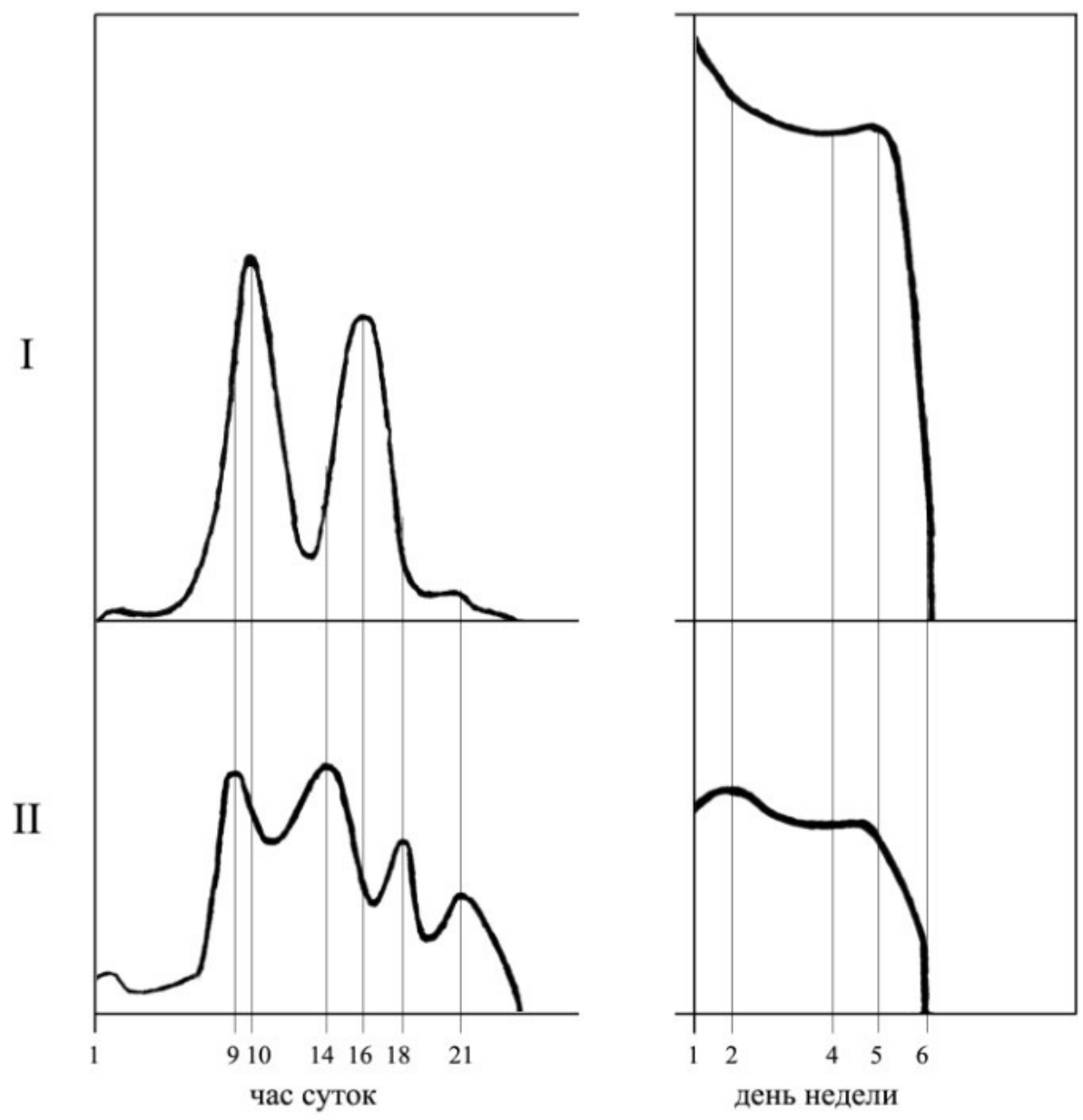

Рис.15. Примерные статистики НС Италии ( I ) и России ( II ).

Приведенные выше результаты исследований позволяют сформулировать наиболее перспективное направление снижения травматизма на предприятии, как включение в систему ежедневного допуск-контроля циркадианных биомедицинских и психофизиологических комплексов процессов, имеющих наиболее высокую степень когерентности с процессом ПТ.

Кроме того, необходим учет временной развертки последействия инструктажей с одновременным согласованием частоты в системе инструктажей с амплитудно-частотными характеристиками ППТ.

## ПРОГРАММА 6 – ВНЕДРЕНИЕ МОДЕЛЕЙ И РЕЗУЛЬТАТОВ ПРИМЕНЕНИЯ КОС

Программа включает 4 проекта.

### Проект 6.1 Внедрение в учебном процессе

Пакет ПО для КОС (**подпроект 2.3.2**) внедрен в учебном процессе ряда высших учебных заведений путем передачи для использования по соответствующим запросам. В частности, методики КОС нашли отражение в методических пособиях для практических и лекционных занятий на радиотехническом и теплоэнергетическом факультетах УГТУ-УПИ, кафедрах атомной энергетики Одесского политехнического института, Московского энергетического института, в учебном пособии Уральского государственного университета. Разработанные методики КОС в части применений в ядерной энергетике и травматологии использованы также в учебном процессе факультета электроники Миланского политехнического института.

### Проект 6.2 КОС в физических измерениях

Методика и фортран-программа распознавания параметров кинетики радиолиза полимерных материалов биологической защиты ЯЭУ внедрена в СФ НИКИЭТ и рекомендована для внедрения на предприятиях Министерства среднего машиностроения. В результате повышена эффективность научных исследований за счет более точного определения констант процессов радиолиза и автоматизации обработки экспериментальных данных.

Библиотека фортран-программ для анализа случайных процессов в составе 38 модулей внедрена в Институте электрофизики УрО РАН для автоматизации расчетов известных, перспективных и новых видов функций, характеристик и параметров случайных процессов и функционирования в составе специального ПО систем автоматизации научных исследований.

**Проект 6.3 Внедрение в АСУ ТП АЭС**

Метод контроля вскипания теплоносителя в активной зоне реактора (**проект 5.2**) внедрен в СФ НИКИЭТ и рекомендован для внедрения на предприятиях Министерства среднего машиностроения. Достигнуто повышение надежности и безопасности эксплуатации реактора.

Методика распознавания режимов работы оборудования, основанная на модели когерентности теплофизических характеристик (**подпроект 5.3.1**) рекомендована для внедрения в программы диагностики информационных каналов, используемых для контроля мощности энергоблока БН-600 Белоярской АЭС им.И.В.Курчатова.

Полученное модельное представление зависимости между изменениями мощности и вакуума в конденсаторах (**подпроект 5.3.2**) вошло в состав Фортран-программы «Диагностика чистоты и эффективности работы конденсаторов турбин ВК-100-6 и К-200-130 на ЭВМ М-4030 с применением методов распознавания образов», внедрено в производство на Белоярской АЭС и передано по запросам для внедрения на Курскую АЭС и Молдавскую ГРЭС, работа рекомендована для внедрения на предприятиях Минэнерго СССР.

**Проект 6.4 Использование в АСУ П предприятия**

Пакет прикладных программ по вычислительной диагностике безопасности и условий труда в СЧМС внедрен в Свердловском филиале Всесоюзного ИПК Минлесбумпрома СССР. Достигнуто сокращение сроков проведения аттестации рабочих мест за счет упрощения и автоматизации измерений и работ, повышена эффективность учебного процесса.

Полученные в ходе моделирования ППТ предложения и рекомендации по форме акта, заменяющей Н-1, внедрены ВЦНИИОТ ВЦСПС для перехода на новую форму акта Н-1 с целью автоматизации анализа травматизма.

На Ясногорском машиностроительном заводе внедрен в опытно-промышленную эксплуатацию головной образец целевой подсистемы Р (предприятия) отраслевой АСУ «Охрана труда», проведено обучение персонала пользованию АСУ «Охрана труда», доложены и переданы для использования в системе инструктажей по технике безопасности результаты исследований последействия инструктажей и синергетических колебаний травматизма в СЧМС.

## ПОЛУЧЕННЫЕ РЕЗУЛЬТАТЫ И ВЫВОДЫ

Выполнен литературно-аналитический обзор по проблематике постановочных задач КОС, ее математического аппарата и приложений к изучению процессов разной физической природы. Реализован 4-х ранговый пакет прототипов и дана их критика.

Сформулирована концепция КОС, как единой системы моделей для анализа процессов и их комплексов, в области частот, наиболее полно отражающей особенности, характерные для СЧМС в целом, и обладающей спецификой методов формирования наблюдений, их обработки и интерпретации. Проведено исследование интегральных, дифференциальных и интегро-дифференциальных признаков и описаний на их основе, методов МО и обоснована перспективность применения ряда ранее не использовавшихся информативных признаков, как для формирования признакового пространства, так и для построения моделей КОС, включающих три и более разнотипных процесса.

Разработан новый метод МО по дисперсии коэффициентов разделяющих функций, дающий устойчивые результаты при гарантированном разделении классов образов исследуемых естественных процессов; усовершенствован метод МО по разрешающей способности использованием идеи случайного поиска с адаптацией.

Предложенные модели применения КОС для изучения невязки, шумов технологических параметров ЯЭУ и оборудования АЭС, процессов ПТ

внедрены в научных исследованиях физических процессов и измерениях констант, при разработке АСУ ТП АЭС, выработке управляющих воздействий для рандомизации и снижения ПТ в практике отделов ОТ и ТБ промышленных предприятий.

Модельные исследования невязки позволили уточнить физические измерения и теоретические представления о процессах, дать более точные рекомендации для проектирования материалов биологической защиты от ионизирующих излучений.

На основе сравнения результатов исследования ППТ России и Италии открыта ранее неизвестная в хронобиологии зависимость между ППТ и циркадианными ритмами, а в более общем плане, установлен детерминированный, полигармонический характер ППТ, с определяющим воздействием биоритмологических циркадианных составляющих и второстепенным – социально-организационных факторов; доказано отсутствие влияния на ППТ околомесячных биоритмов, широко применяемых на предприятиях ряда стран; сформулирована автоколебательная модель ППТ и предложены методы его рандомизации для снижения уровня травматизма на производстве.

В области обеспечения исследований и технологических применений разработан и внедрен пакет программ КОС, включающий все блоки ПО от формирования исходных данных до получения аналитических результатов.

Общие выводы: поставленная цель – развить систему моделей и методов КОС интеграцией с распознаванием образов на всех этапах получения информации об исследуемых процессах и СЧМС – достигнута; разработаны контуры нового научного направления в комплексной обработке сигналов процессов различной физической природы.

## ОСНОВНЫЕ ПУБЛИКАЦИИ